\documentclass[12pt]{article}
\pdfoutput=1

\usepackage{epsfig,a4wide,amsmath,amssymb,amsfonts,mathrsfs,amsthm}
\usepackage{mathtools}
\usepackage{xcolor,yfonts}
\usepackage{graphicx}
\usepackage{subcaption}
\usepackage[utf8]{inputenc}
\usepackage{siunitx}
\usepackage{rotating}
\usepackage{hyperref}

\usepackage[normalem]{ulem}
\usepackage{cite}
\usepackage{float}
\usepackage{soul}
\usepackage{amsmath}
\usepackage{amssymb}
\usepackage{color}
\usepackage{geometry}
\usepackage{comment}

\numberwithin{equation}{section}

\newcommand{\dd}{\mathrm{d}}
\newcommand{\DD}{\Delta\!\!\!\!\Delta}
\newcommand{\Grad}{\nabla\!\!\!\!\nabla}

\begin{document}

\begin{titlepage}
\begin{center}

~\\[1cm]

{\Large \bf 
Are $S^1\times S^2$ wormholes generic with large sources?
}

~\\[0.5cm]

{\fontsize{14pt}{0pt} 
Xiaoyi~Liu${\ }^{\diamond}{\ }^{\dagger}$,  
Donald~Marolf${\ }^{\diamond}$, 
Jorge~E.~Santos${\ }^{\star}$}

~\\[0.1cm]

\it{ ${}^{\diamond}$  Department of Physics, University of California, Santa Barbara, CA 93106, USA}

~\\[0.05cm]

\it{ ${}^{\dagger}$ Perimeter Institute for Theoretical Physics,}\\
{31 Caroline Street North, Waterloo, Ontario N2L 2Y5, Canada }

~\\[0.05cm]

\it{ ${}^{\star}$ Department  of  Applied  Mathematics  and  Theoretical  Physics,  }\\
\it{University  of  Cambridge, Wilberforce Road, Cambridge, CB3 0WA, UK}
\end{center}

\vspace{30pt}
\noindent
Euclidean path integrals can be used to prepare states of a Lorentzian QFT.  So long as any sources are turned off on the $t=0$ surface, the resulting Lorentzian states all belong to the same Hilbert space.  Constructing more states than allowed by the Lorentzian density of states means that the resulting states must be linearly dependent.  For large amplitude sources and a fixed cutoff on energy, the AdS bulk dual of this effect has been conjectured to be captured by spacetime wormholes.  Wormholes should then be generic in the presence of large such Euclidean sources.

This hypothesis can be studied in a context with asymptotically locally AdS$_4$ boundaries of topology $S^1 \times S^2$ in which the wormhole is supported by a source for minimally-coupled massless bulk scalars.  In preparation for a later more complete study, we consider here a preliminary toy version of the model in which the spacetimes are cohomogeneity-1, but with the consequence that the sources do not vanish at $t=0$.   
We then find that generic sources at large masses do {\it not} lead to wormholes. Along the way we map out the phase diagram for wormhole, thermal AdS,
and black hole phases of our cohomogeneity-1 ansatz. We also numerically evaluate their stability by identifying negative modes.  In parallel with the previously-studied case of $S^3$ boundaries, the results are analogous to those associated with the familiar Hawking-Page transition.  

\vfill
\noindent
\end{titlepage}

\newpage

\tableofcontents
\baselineskip=16pt

\section{Introduction}
Euclidean spacetime wormholes have been of great interest in recent studies of quantum gravity.  In particular, they give rise \cite{Coleman:1988cy,Giddings:1988cx,Penington:2019kki,MarolfMM:2020xie} to effects  associated with resolving the black hole information puzzle (see also \cite{Almheiri:2019qdq}) and, in contexts where the dominant contribution is associated with black holes (e.g., with asymptotically AdS boundary conditions), with ensuring that the density of states is of size $e^{S_{\rm BH}}$ (where $S_{\rm BH}$ is the Bekenstein-Hawking entropy); see also \cite{Saad:2018bqo,Saad:2019lba,Balasubramanian:2022gmo,Balasubramanian:2022lnw}.   
 A primary such mechanism is that wormhole effects cause the norms of certain otherwise-nontrivial states to vanish\footnote{At least in each baby universe superselection sector \cite{Coleman:1988cy,Giddings:1988cx,MarolfMM:2020xie}.}, so that one finds new linear relations (and a correspondingly smaller Hilbert space dimension) than would be the case without such wormholes.

As emphasized in \cite{Balasubramanian:2022gmo,Balasubramanian:2022lnw}, it is particularly interesting to study states created by applying operators of the form $e^{-\beta H}A_\beta$, where the factor $A_\beta$ is an operator whose `strength' is chosen to increase with $\beta$ in the sense that one holds fixed the total energy of the resulting state; see also \cite{Antonini:2023hdh}.  In terms of a gravitational path integral, at large $\beta$ the factor $e^{-\beta H}$ is associated with taking the length of the Euclidean boundary to be long, and the factor $A_\beta$ is then associated with some deformation of the boundary conditions in the distant Euclidean past.  In the context of AdS/CFT, the insertion of this $A_\beta$ corresponds to the activation of a certain source for the dual CFT.  We are then interested in a limit where $\beta \rightarrow \infty$ while the source becomes large in such a way that the energy remains fixed. 

\begin{figure}
    \centering
    \includegraphics[width=0.6\linewidth]{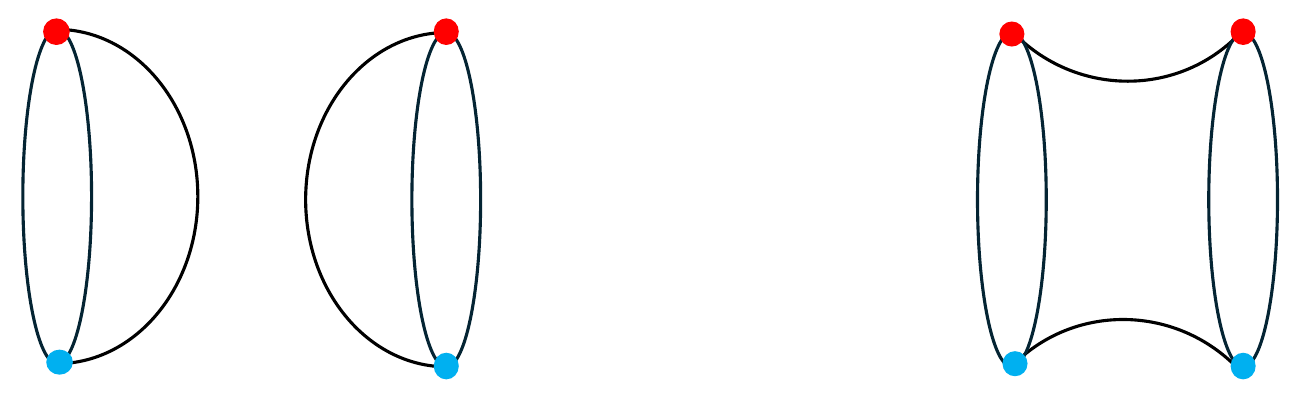}
    \caption{Possible contributions to bulk path integrals computing (products or moments of) inner products of dual CFT states defined by insertions of operators of the form $e^{-\beta H}A_{\rm red}$ with those defined by insertions $e^{-\beta H}A_{\rm blue}$.  Insertions of $A_{\rm red}, A_{\rm blue}$ are shown respectively as red/blue dots.  The disconnected solution at left generally has large Euclidean action because these fields are distinct in the bulk low-energy theory, so that the bulk inner product is small at this EFT level.   The wormhole geometry is thus expected to have lower action. But the Euclidean action of the connected contribution at right need not become large in this limit and thus may be expected to dominate.  }
    \label{fig:nhalfwh}
\end{figure}

Each such state on its own is then naturally expected to lead to a Lorentzian black hole\footnote{So long as black holes dominate the microcanonical ensemble at the given energy.  With AdS boundary conditions, this condition will automatically hold in the limit $G \rightarrow 0$ unless the energy is chosen to vanish in that limit.} with fixed total energy.  However, the large value of $\beta$ and the strength of the operator $A_\beta$ suggest that the black hole will have a large and, in some sense, highly excited interior.  As a result, if one neglects wormhole effects and computes semiclassically, one should find such states to be linearly independent.  But since the total number of states at fixed mass should be bounded by $e^{S_{\rm BH}}$,  and since $\beta$ can be arbitrarily large, there must be important corrections to this result.  One thus expects wormhole effects to dominate computations of inner products of such states; see figure \ref{fig:nhalfwh}.  In particular, \cite{Balasubramanian:2022gmo} studied asymptotically locally AdS (AlAdS) on-shell Euclidean wormholes supported by spherical shells of pressureless dust and with boundary topology $S^1 \times S^2$ and suggested that,  for generic operators in the limit of large $\beta$,  the wormhole effects expected from the above insertions of $e^{-\beta H} A_\beta$ should be similar.

This conjecture is particularly interesting in the case where $A_\beta$ is defined by choosing a nontrivial boundary condition for light bulk fields (which we take to be massless scalars). An important feature of this setup is that, although the wormholes studied in \cite{Balasubramanian:2022gmo} enjoy an exact $\mathrm{SO}(3)$ symmetry that acts as rotations of the $S^2$, on-shell wormholes with this symmetry are now strictly forbidden (regardless of whether one breaks the corresponding $\mathrm{U}(1)$ rotational symmetry of the $S^1$).  The prohibition can be seen as a consequence of topological censorship \cite{Friedman:1993ty}, since Wick rotating the $S^2$ to $\mathrm{dS}_2$ would yield a traversable wormhole with de Sitter cross-sections in a classical theory satisfying the null energy condition. Here traversability follows immediately from the fact that $\mathrm{dS}_2$ has a static patch, so that there is an infinite amount of static patch time for any signal to travel from one boundary to the other.   In contrast, when the sources vary over the sphere, the dependence on the associated angles means that Wick-rotating any angle on the sphere leads to bulk fields that are intrinsically complex and, as a result, violate the null energy condition.  Such violations then allow wormholes to exist.  In order for on-shell wormholes to dominate generically at large sources, it must thus be the case that, in the presence of a large $\mathrm{SO}(3)$-symmetric source, only a tiny amount of symmetry breaking would be required to form a wormhole\footnote{An identical argument also shows wormholes to be forbidden when one preserves the $\mathrm{U}(1)$ symmetry of the $S^1$, even if one then breaks the rotational symmetry on the $S^2$.  In general, all geometric  symmetries must be broken for a Euclidean wormhole to exist when Wick rotation preserves the boundary conditions and leads to a Lorentzian theory that satisfies the null energy condition.}.

\begin{figure}
    \centering
    \includegraphics[width=0.9\linewidth]{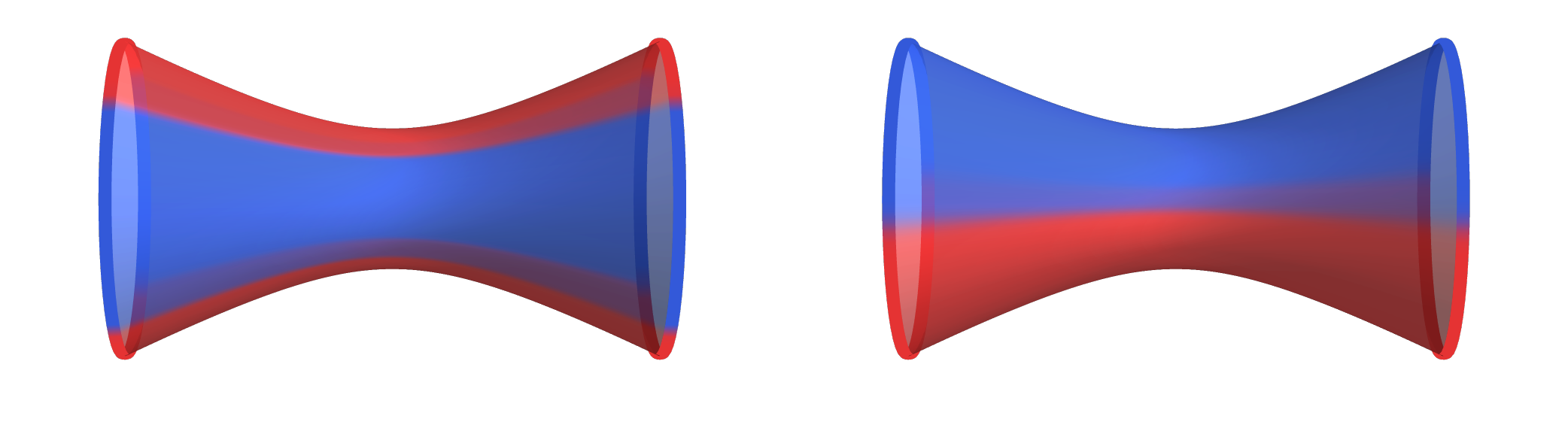}
    \caption{Wormholes with $S^1$ boundaries with shading indicating the rotating phase of a complex scalar $\phi$. The $S^2$ factor of our setup is not shown.  Blue shading indicates regions where $|{\rm Re} \phi| > |{\rm Im} \phi|$, while red indicates $|{\rm Re} \phi| < |{\rm Im} \phi|$.  The left panel thus shows a case with angular momentum $n=1$ on the $S^1$ while the right panel shows $n=1/2$ (which arises if we allow antiperiodic scalars). That such boundary conditions model the insertion of $e^{-\beta H}A_{\mathrm{red/blue}}$ in a dual CFT can be seen by noting that the $n=1/2$ case may be interpreted as a less well-localized version of the right panel of figure \ref{fig:nhalfwh2}.}
    \label{fig:nhalfwh2}
\end{figure}

Here we take a first step toward investigating the above conjecture.  We consider a toy model defined by imposing an ansatz so that the bulk AlAdS geometries have cohomogeneity-1.  In particular, by including a sufficient number of complex scalars we may choose the scalar sources to completely break the rotational symmetries of both the $S^1$ and $S^2$ factors while preserving a different `diagonal' $\mathrm{U}(1) \times \mathrm{SO}(3)$ symmetry that acts as a rotation on the space of scalar fields (with the $\mathrm{U}(1)$ rotating the scalars by a phase) while simultaneously rotating the $S^1$ and $S^2$ factors of the geometry. As illustrated in figure \ref{fig:nhalfwh2}, such boundary conditions take the general form associated with computing the inner product of states that might naively have been expected to be distinct in the limit of large scalar sources; note the analogy with figure \ref{fig:nhalfwh}.  

However, since the bulk stress tensor and geometry transform trivially under rotations of the space of scalars, the above diagonal symmetry in fact makes these two quantities invariant under a purely geometric $\mathrm{U}(1) \times \mathrm{SO}(3)$. The bulk geometry thus depends only on a single radial coordinate $r$; i.e., it is cohomogeneity-1.

We find it convenient to fix boundary conditions such that a $2\pi$ rotation of the $\mathrm{U}(1)$ factor always corresponds to rotating the phase of our scalars by precisely $2\pi$.   Such boundary conditions thus have angular momentum $n=1$ on the $S^1$.    We can then add such an $n=1$ source that is nevertheless symmetric on the $S^2$ (i.e., which involves only the $\ell=0$ spherical harmonic on $S^2$) with some large amplitude $V_0$ and ask what amplitude $V_1$ for an $n=1,\ell=1$ source is required in order for wormholes to exist.    Our main result is numerical evidence that the required $V_1$ remains finite in the limit of large $V_0$, indicating that on-shell wormholes do {\it not} in fact dominate generically in that limit.    

However, a consequence of our symmetry-preserving ansatz is that our sources are non-zero at all points on our AlAdS boundaries.  As a result, the states we prepare do not all live in the same Lorentzian theory.  This is an important distinction from the setting considered in \cite{Balasubramanian:2022gmo,Antonini:2023hdh} that we hope to remove in future work. 

Along the way, we map out the general phase diagram for our scalar-supported $S^1 \times S^2$ wormholes and study their stability (existence of negative modes).   With two asymptotic boundaries we find three families of solutions: wormholes, disconnected thermal AdS solutions, and disconnected Euclidean black holes, each of which includes back-reaction from bulk the scalar fields.  As for the quantum-corrected Jackiw-Teitelboim wormholes described in \cite{Garcia-Garcia:2020ttf}, the constrained wormholes studied in \cite{Cotler:2020lxj}, and the spherical wormholes previously explored in \cite{Marolf:2021kjc}, the wormhole phase appears only for sufficiently large scalar sources, and it dominates only at even larger source amplitudes (above a Hawking-Page-like transition).  Our studies may also be of interest in the context of constructing field-theoretic versions of other wormholes supported by particles and/or dust (see e.g. \cite{Belin:2020hea,Belin:2021ryy,Chandra:2022bqq,Chandra:2022fwi,Wang:2025bcx}), perhaps in connection with exploring configurations with higher cohomogeneity (in which a particle description is inherently singular) or with regard to exploring stability of various constructions of Euclidean spacetime wormhole (for which we find no guidance in the literature for situations involving continuous distributions of dust).

To provide some orientation for the reader, we begin in section \ref{sec:FLRW} with some basic observations regarding what is required to support a Euclidean wormhole with homogeneous cross-sections of the form $S^1 \times S^2$ (and for slightly more general cross sections as well).  We then present our model and the phase diagram for $\ell >0$  sources in section \ref{sec:theoryandPD}.   Section \ref{sec:ell0} studies our model with an additional $\ell=0$ source and presents our main results concerning non-genericity of wormholes with large such sources.  We also study stability of our wormholes in section \ref{sec:stability} using a method based on \cite{Kol:2006ga} (which is closely related to the methods of \cite{Garriga:1997wz,Gratton:1999ya}).  Here we take care to explain certain subtleties of such computations that may appear to lead to non-physical divergences related to those recently discussed in \cite{Marolf:2025evo}.  We also check in appendix \ref{app:ROT} that the rather different `rule-of-thumb' approach of \cite{Marolf:2022ntb} yields equivalent results when applied using the DeWitt$_{-1}$ metric on the space of perturbations.  We close with final comments in section \ref{sec:concl} summarizing the results and discussing future directions. 

\section{FLRW-like approach}\label{sec:FLRW}

Here we provide some brief comments motivating expectations for wormholes with boundary topology $S^1 \times K$ with cohomogeneity-1 and AdS asymptotics.
The metric of a $(3+1)$-dimensional such wormhole may be written in a Euclidean FLRW-like form,
\begin{equation}
\dd s^2=\dd r^2+b(r)^2\dd \tau^2+a(r)^2\dd\Sigma_{2}^2\,,
\end{equation}
where $\dd\Sigma_2^2$ is the metric on the $2$ dimensional sphere $\mathbb{S}^{2}$ ($k=1$), Euclidean space $\mathbb{R}^{2}$ ($k=0$), hyperbolic space $\mathbb{H}^{2}$ ($k=-1$), or a quotient $K$ of one of these spaces, and where $\tau$ is perioically identified.  The $rr$ component of the Einstein equation \begin{equation}
    R_{ab}-\frac{1}{2}R g_{ab}-\frac{3}{L^2}g_{ab}=8\pi G T_{ab},
\end{equation}
is then a Friedman-like equation which, in the presence of a negative cosmological constant
with AdS length $L$, takes the form
\begin{equation}\label{eq:Friedmanneq}
    \left(\frac{\dot{a}}{a}\right)^2+\frac{2\dot{b}}{b}\cdot\frac{\dot{a}}{a}=
    -16\pi G\rho
     +\frac{k}{a^2}+\frac{3}{L^2}\,,
\end{equation}
where $T_{ab}$ is the stress energy tensor of the bulk matter, and $\rho=-T_{rr}$ plays the role of a Lorentz-signature energy density (i.e., our negative cosmological constant has the same effect as a constant negative $\rho$). Here a dot ($\cdot$) denotes a derivative with respect $r$, which here plays the role of a Euclidean ``time". At conformal infinity $(r\to\pm\infty)$\,, we require $a(r)\to\infty$ to satisfy the asymptotic AdS boundary conditions. Any wormhole will thus have some $r=r_*$ where $a$ takes its minimum $a(r_*)=a_0$ and where $\dot a|_{r=r_*}$ vanishes. The left-hand side of equation \eqref{eq:Friedmanneq} clearly vanishes at $r_*$, so  the right-hand side must vanish as well. Since the last term $3/L^2$ on the right-hand side is positive definite, we see that there must be some compensating negative contribution from the first two terms.  For example, $3$-dimensional wormholes can be supported by conical defects \cite{Chandra:2022bqq}\ since such defects  contribute positively to $\rho$.

However, the present work focuses on contexts where the matter sources described by a standard field theory.  There are then 4 possible ways to obtain positive contributions to the right-hand-side of \eqref{eq:Friedmanneq}.  The first is to take $k=-1$ so that our wormhole has boundary topology $S^1\times \mathbb{H}^{2}$ (or a quotient thereof).  There can then be wormholes analogous to those studied by Maldacena and Maoz \cite{Maldacena:2004rf}.

The other options describe various ways to obtain $\rho >0$.  One choice is to choose fields engineered so that variations with $r$ provide a `kinetic energy' that contributes positively to $\rho$.  However, for a standard field this kinetic energy is actually negative.  This can be seen from the fact that, in Lorentz signature, the analogous kinetic energy would contribute with a positive sign.  But Wick-rotating to Euclidean time changes the sign of the time-time component of the metric, and thus the sign of this kinetic energy contribution to $\rho$.  As a result, the desired positive contribution is obtained only for fields with a `wrong-sign' kinetic energy term such as that found in standard treatments of Euclidean axions.  The canonical examples of this approach are thus the axion wormholes of e.g. \cite{Giddings:1987cg,Giddings:1989bq} (and their cousins \cite{Arkani-Hamed:2007cpn,Andriolo:2022rxc,Jonas:2023ipa,Gutperle:2002km,Aguilar-Gutierrez:2023ril} which include dilatons and other choices of asymptotics).

Another choice appears to simply be to give some scalar field a positive potential that turns off as $r\rightarrow \pm \infty$.   This choice appears not to have been pursued in the literature but might be interesting to consider, e.g. for a double-well potential with minimum $V=0$ and boundary conditions that require a domain wall in the bulk.   However, the potential would need to be arranged so that the gradient of the scalar is small inside the wall, as such gradients suppress wormholes when the scalar field kinetic term has the standard sign.  

The final option is to use the term in $\rho$ that arises from gradients of the matter fields on the $S^1$ or on the two-dimensional compact factor $K$.   Wormholes supported in this way were first investigated in \cite{Marolf:2021kjc} for the case of $S^{3}$  cross sections (i.e., with no $S^1$ factor), and we will generalize such solutions to boundary topology $S^1 \times S^2$ below.  We remind the reader that, as noted in the introduction, topological censorship implies that such wormholes solutions can exist only such gradients break all $\mathrm{U}(1)$ symmetries.  In particular, we can find such wormholes only when we activate gradients along both the $S^1$ and $S^2$ factors of the geometry.

\section{$S^1 \times S^2$ wormholes in a gravity theory with scalars}
\label{sec:theoryandPD}

We present the details of our model and our cohomogeneity-1 ansatz in section \ref{sec:EGCS} below.  We then present results for the associated wormholes, black holes, and thermal AdS solutions in the following subsections.  In particular, we construct the phase diagram showing which class of solution has the smallest action in each region of parameter space.  The full phase diagram is displayed in section \ref{sec:PD1}.

\subsection{Cohomoegneity-1 $S^1\times S^2$ Einstein gravity with complex scalars}
\label{sec:EGCS}

As remarked above, we focus in this work on Euclidean-signature  asymptotically locally AdS solutions with cohomogeneity-1 and boundary topology $S^1 \times S^2$.  The fact that our solutions have cohomogeneity-1 means that the components of both the metric and the scalar field stress tensor can be taken to depend only on a single coordinate $r$.
 
To construct such solutions with non-negative integer angular momentum $\ell$ on the $S^2$, we include $(2\ell+1)$ complex minimally-coupled massless scalar fields. We will choose the scalar fields to have a special angular dependence such that they are invariant under a diagonal action of $\mathrm{SO}(3)$ that simultaneously rotates both the $S^2$ factor of the geometry and the space of scalar fields.  We similarly take our ansatz to be invariant under the diagonal $\mathrm{U}(1)$ that simultaneously rotates the $\mathrm{U}(1)$ factor of the geometry and the phase of our complex scalars. Since the stress tensor is separately invariant under both phase rotations of our complex scalars and rotations of the space of scalar fields, it will be invariant under purely geometric rotations of the $S^1$ and $S^2$ factors of the spacetime.  We may thus take the metric to be rotationally invariant as well.
 
The $(2\ell+1)$ scalar fields can be viewed as components of a vector $\vec{\Phi}$:
\begin{equation}
(\vec{\Phi})_m=\phi_m\,,
\end{equation}
with $|m|\leq \ell$. With an appropriate normalization of the scalar fields, the action may be written in the form
\begin{equation}
\label{eq:action1}
S=-\frac{1}{16\pi G}\int_{\mathcal{M}}\mathrm{d}^4x\sqrt{g}\left(R+\frac{6}{L^2}-2\nabla^a \vec{\Phi} \cdot \nabla_a\vec{\Phi}^\star\right)-\frac{1}{8\pi G}\int_{\partial \mathcal{M}}\mathrm{d}^3x\sqrt{h}K-S_{\partial{\mathcal{M}}}\,.
\end{equation}
Here ${}^\star$ denotes complex conjugation, $\cdot$ denotes the usual Cartesian dot product between vectors in $\mathbb{R}^{2\ell+1}$, and $L$ is the AdS length scale.  The explicit integral over $\partial {\cal M}$ is the usual Gibbons-Hawking-York term, where $h$ is the determinant of the induced metric on $\partial\mathcal{M}$ and $K$ is the trace of extrinsic curvature associated with an outward-pointing normal on the same hypersurface. In addition, $S_{\partial{\mathcal{M}}}$ is a sum of boundary counter-terms chosen to make the action finite and the variational problem well-defined. We work in the grand canonical ensemble, where we fix the boundary induced metric and the values of $\vec{\Phi}$ at the boundary (i.e, in the language of AdS/CFT we fix the dimension-zero source dual to the operator dual to $\vec{\Phi}$). 

In this ensemble, using Greek indices to denote boundary coordinates, the counter terms take the form
\begin{equation}
S_{\partial{\mathcal{M}}}=-\frac{1}{4\pi GL}\int_{\partial \mathcal{M}}\mathrm{d}^3x \sqrt{h}-\frac{L}{16\pi G}\int_{\partial \mathcal{M}}\mathrm{d}^3x \sqrt{h}R^h+\frac{L}{8\pi G} \int_{\partial \mathcal{M}}\mathrm{d}^3x \sqrt{h}h^{\mu\nu}\nabla^h_{\mu}\vec{\Phi}\cdot \nabla^h_{\nu}\vec{\Phi}^\star\,,
\end{equation}
where $R^h$ is the Ricci scalar of the induced boundary metric $h_{\mu\nu}$ and $\nabla^h$ its metric-compatible covariant derivative. We define the action \eqref{eq:action1} as the $\epsilon \rightarrow 0$ limit of the action for a version of the spacetime that is cut off at the surface 
$z=\epsilon\to 0$ as defined by the standard Fefherman-Graham expansion:
\begin{equation}
    \dd s^2=\frac{L^2}{z^2}\left\{\dd z^2+\left[g^{(0)}_{\mu\nu}(x)+\mathcal{O}(z^2)\right]\dd x^\mu\dd x^\nu\right\}\,,
\end{equation}
such that 
\begin{equation}
    \lim_{\epsilon\to 0}\epsilon^2 h_{\mu\nu}=L^2 g^{[0]}_{\mu\nu}\,
\end{equation}
for some specified boundary metric $g^{[0]}_{\mu\nu}$.  Below, we will focus on boundary metrics of the form
\begin{equation}
\mathrm{d}s^2_{\partial \mathcal{M}} = g^{[0]}_{\mu\nu} dx^\mu dx^\nu=\mathrm{d}\tau^2+L^2 \mathrm{d}\Omega_2^2\,,\quad \mathrm{d}\Omega_2^2=\mathrm{d}\theta^2+\sin^2\theta\mathrm{d}\varphi^2\,,
\end{equation}
with $\theta\in[0,\pi]$, $\varphi\sim\varphi+2\pi$ and $\tau\sim\tau+\beta\,L$.
Since the physics is invariant under conformal (Weyl) rescalings of the boundary metric, we consider only boundary metrics in which the sphere has unit radius as above.  Note that by definition, $\beta$ is dimensionless.

As described earlier, we would like the scalar fields to carry angular momentum $\ell$ on the $S^2$ and angular momentum $n \in {\mathbb Z}$ on the $S^1$ while maintaining $\mathrm{U}(1)\times \mathrm{SO}(3)$ invariance of the stress tensor.  This can be achieved by following the treatment of an analogous problem \cite{Marolf:2021kjc} and imposing the ansatz
\begin{equation}
\label{eq:phiansatz}
\phi_m(\tau,r,\theta,\varphi) = \phi(r)e^{-i \frac{2\pi n}{\beta\,L} \tau}\times \sqrt{\frac{(\ell-m)!}{(\ell+m)!}}\times P^{\ell}_m(\cos \theta)\times\left\{
\begin{array}{ll}
\sqrt{2}\,\sin (m\,\varphi)\,,& \text{for}\quad m<0
\\
1\,, & \text{for}\quad m=0
\\
\sqrt{2}\,\cos (m\,\varphi)\,,& \text{for}\quad m>0
\end{array}\right.\, ,
\end{equation}
 where $P^{\ell}_m(x)$ are the standard associated Legendre polynomials.  Since $n$ is an integer, the functions $\phi_m(\tau,r,\theta,\phi)$ satisfy the relevant periodic boundary conditions along the thermal-$\tau$ circle. The normalizations are chosen so that 
 \begin{equation}
 \sum_{m=-\ell}^{\ell} |\phi_m(\tau,r,\theta,\varphi)|^2=\phi(r)^2\,.
 \end{equation}
 Since the scalar field is massless, the source $V$ of the operator dual to $\Phi$ in any dual CFT is simply
 \begin{equation}
V:=  \lim_{r\to\pm\infty}\phi(r)\,.
 \end{equation}
 Here $r\rightarrow\pm \infty$ corresponds respectively to the right and left conformal boundaries of our asymptotically locally AdS wormhole.
 
The resulting solution is then specified by the parameters $(n,\ell,V,\beta)$. Note, however, that since the geometry is invariant under rotations of the $S^1$ factor, given any such solution we may construct new `solutions' simply by first unrolling the $S^1$ into a real line $\mathbb R$ and then identifying the result with an arbitrary period (which, for the moment, we allow to perhaps make the scalar fields multi-valued on the resulting $S^1$).  The new solution then has parameters $(n',\ell,V,\beta')$ with the same wavelength $\beta'/n' = \beta/n$ as the original one.  We thus see that the essential properties of our wormholes depend only on the ratio $\beta/n$.  Unless explicitly noted otherwise, We will therefore simply fix $n=1$ in all further discussions below.

We will proceed by constructing solutions numerically.    Specifically, we take the metric to be of the form
 \begin{equation}
 \label{eq:gansatz}
 \mathrm{d}s^2=\frac{\mathrm{d}r^2}{g(r)}+f(r)\mathrm{d}\tau^2+p(r)\mathrm{d}\Omega_2^2\, ,
 \end{equation}
where the range of $r$ will be specified below.   At this stage we have not fixed any particular definition radial coordinate $r$.  Indeed, it will be convenient to make different choices for different classes of solutions. For black holes we expect $f$ and $g$ to have a simple zero at some finite value of $r$ that corresponds to the radius of the black hole horizon, and we expect $p(r)$ to be positive throughout. For the thermal AdS solutions, we expect $p(r)$ to vanish at the origin. As a result, for both of these classes of solutions we will impose the Schawrzschild gauge $p(r)=r^2$.   For black holes we choose some positive $r_+$ and require $r\ge r_+$ with $f(r_+)=g(r_+)=0$, while for thermal AdS type solutions we take $r\ge0$ with $g(0)=1$.  For wormhole solutions we will take $r$ to run over the entire real line (including both negative and positive values).  

However, the wormholes of interest will have the same sources on both the right and left boundaries.  We thus expect only solutions that are invariant under a ${\mathbb Z}_2$ reflection.  As a result, it suffices to construct only the right half of the wormhole. In practice we thus restrict to $r\ge 0$ and use the gauge $p(r)=r^2+r_0^2$ with the Neumann boundary condition $f^\prime(0)=g^\prime(0)=0$ at the wormhole neck, $r=0$.   We similarly note that taking two copies of either the thermal AdS or black hole solutions give disconnected bulk geometries with a pair of $S^1\times S^2$ boundaries (with one connected boundary for each connected component of the bulk).

The equations of motion that follow from our action take the form \begin{subequations}
\begin{equation}\label{eq:EinsteinEq}
R_{ab}-\frac{R}{2}g_{ab}-\frac{3}{L^2}g_{ab}=\nabla_a \vec{\Phi}\cdot\nabla_b \vec{\Phi}^\star+\nabla_b \vec{\Phi}\cdot\nabla_a \vec{\Phi}^\star-g_{ab}\nabla_c \vec{\Phi}\cdot\nabla^c \vec{\Phi}^\star\,,
\end{equation}
and
\begin{equation}
\nabla_a \nabla^a \vec{\Phi}=0\,.
\end{equation}
\end{subequations}
Inserting the ansatz \eqref{eq:gansatz} yields
\begin{subequations}\label{eq:eomsr}
\begin{equation}
\left[\sqrt{f(r)} \sqrt{g(r)} p(r) \Phi^\prime(r)\right]^\prime-\frac{\ell  (1+\ell ) f(r)+\omega ^2 p(r)}{\sqrt{f(r)} \sqrt{g(r)}} \Phi (r)=0\,,
\end{equation}
\begin{multline}
\frac{g^\prime(r) p^\prime(r)}{p(r)}-\frac{2 \omega ^2 \Phi (r)^2}{f(r)}-\frac{6}{L^2}-\frac{g(r) p^\prime(r)^2}{2 p(r)^2}+2 g(r) \Phi^\prime(r)^2
\\
-\frac{2}{p(r)}+\frac{2 \ell  (\ell+1) \Phi(r)^2}{p(r)}+\frac{2 g(r)
   p^{\prime\prime}(r)}{p(r)}=0\,,
\end{multline}
\begin{equation}
\frac{f^\prime(r) p^\prime(r)}{2 f(r) p(r)}-\frac{3}{L^2 g(r)}-\frac{1}{g(r) p(r)}+\frac{\omega ^2 \Phi (r)^2}{f(r) g(r)}+\frac{\ell  (\ell +1) \Phi (r)^2}{g(r) p(r)}+\frac{p^\prime(r)^2}{4 p(r)^2}-\Phi^\prime(r)^2=0\,.
\end{equation}
\end{subequations}
Here $\omega\equiv 2\pi n/(\beta L)$. Further details of the actual construction of these solutions are given in Appendix \ref{App:construction}.

\subsection{Phases and dominance}
\label{sec:phasesbeta1}
\label{sec:PD1}

As noted above, it suffices to consider only the case $n=1$.  To begin our discussion of phases and their relative dominance, let us momentarily fix the conformally-rescaled length $\beta$ of the boundary $\tau$-circle at the conformal infinity by fixing $\beta=1$ and exploring which classes of solutions exist for various values of $V$ and $\ell$\,. 

\begin{figure}[h!]
    \centering
    \includegraphics[width=0.5\linewidth]{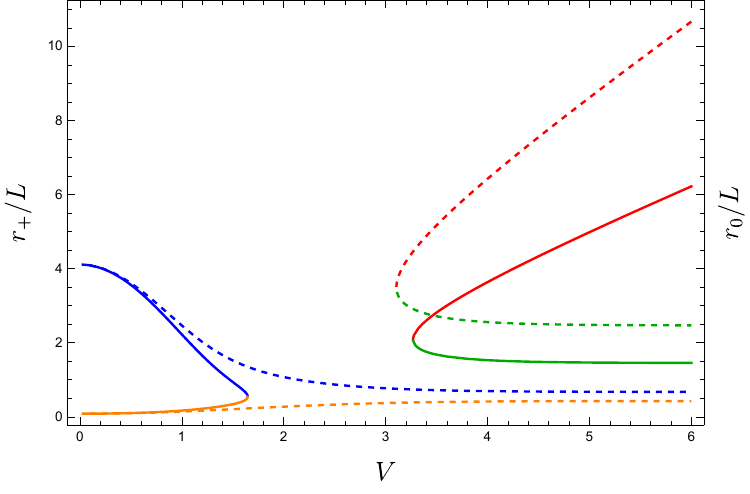}
    \caption{The minimum radius ($r_+$ or $r_0$) of the $S^2$ for the black holes and wormholes for $\beta=1$ with either $\ell=1$ (solid lines) or $\ell=2$ (dashed lines). Blue/orange describes the large/small black hole branches, while red/green shows data for the large/small wormholes.  
}
    \label{fig:fxibeta}
\end{figure}

Our results are shown in figure \ref{fig:fxibeta}.
At fixed $\ell$, we find wormholes only when the boundary scalar source $V$ exceeds an $\ell$-dependent critical value $V_{\rm W}$. 
This threshold effect is expected from our FLRW-like analysis in section \ref{sec:FLRW}.  In particular, 
since our $S^2$ has $k=+1$, only for large enough boundary sources do we expect the induced spatial gradients of our scalars to make $-\rho$  sufficiently negative so as to cancel the positive contributions from the last two terms on the RHS of the Friedmann-like equation \ref{eq:Friedmanneq}.

Turning now to the black holes, when $\ell=1$ we find that black holes exist only for $V$ {\it below} a critical value $V_{\rm BH} \approx 1.6482(8)$ at which the large and small black holes merge.  However, for $\ell \ge 2$ and for $\beta=1$, if we looked hard enough we have always found both large and small black hole solutions with distinct values of $r_+$ (though at fixed $\beta$ the solutions become more difficult to find as we increase $V$). This appears to be due to the fact that, while scalar sources generally increase the temperature required to nucleate black holes (and also for the Hawking-Page transition at which black holes dominate over thermal AdS), for large enough $\ell$ we may ignore the positive curvature of the $S^2$ and one thus expects behavior analogous to what one would find if one replaced the $S^2$ by a flat torus (for which the nucleation of black holes occurs already at $T=0$).

Indeed, just as in the case with no scalars ($V=0$), in the interior of the regime where black holes exist we in fact find two black hole solutions with different values of $r_+$.  We refer to these solutions as the large/small black holes according to the relative sizes of $r_+$.  Similarly, when $V$ exceeds the threshold for the existence of wormholes, we in fact find two wormhole solutions with differing values of $r_0$. We call these the large/small wormholes according to the relative sizes of $r_0$.
For $\ell=1$ the large/small wormholes join at $V_{\rm W}\approx 3.2705(8)$, while this occurs at 
$V_{\rm W}\approx 3.1067(6)$ for $\ell=2$.

\begin{figure}[h!]
    \centering
    \includegraphics[width=\linewidth]{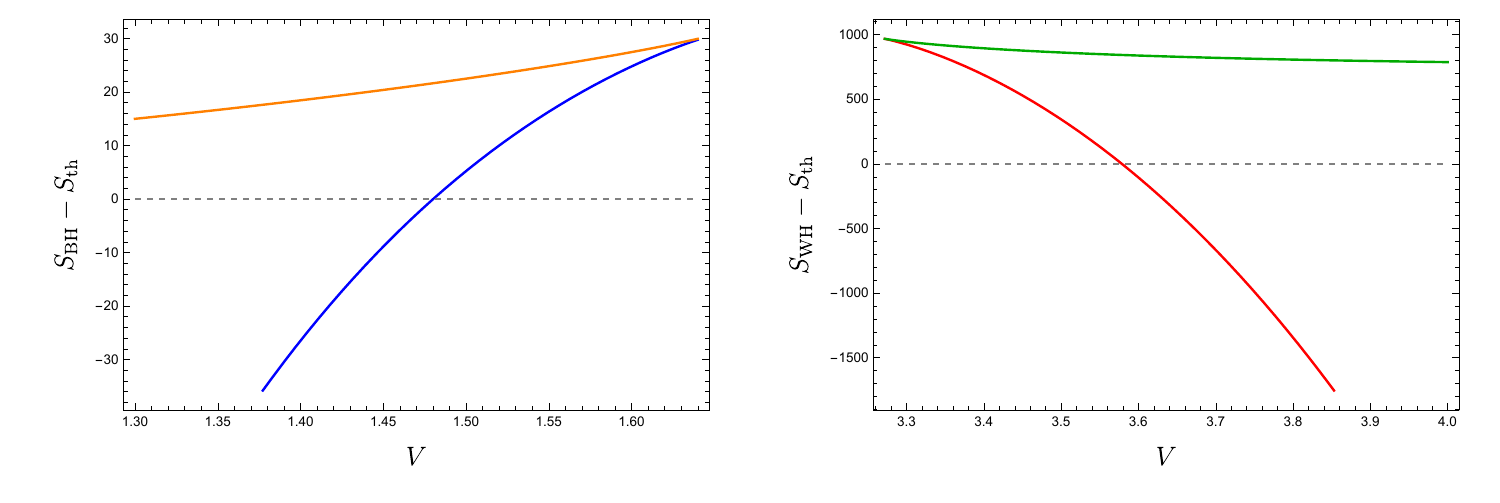}
    \caption{Comparison of the actions for black holes (left) or wormholes (right) with that of thermal AdS for $\ell=1\,,\beta=1$.  $S_{\rm BH}$ and $S_{\rm th}$ are the actions for a {\it single} copy of black hole or thermal AdS geometry, see  \eqref{eq:bhS} and \eqref{eq:thS}, while $S_{\rm WH}$ is half of the action for the 2-boundary wormhole.
    Phase transitions occur when the colored curves cross the dashed gray line. As in figure~\ref{fig:fxibeta}, orange/blue and green/red denote small/large black holes or wormholes.   }
    \label{fig:actiondiffell1fixbeta}
\end{figure}

One expects the solution with lowest Euclidean action to dominate the ensemble at each set of parameters. Results for the actions of the various branches of solutions are shown in figure \ref{fig:actiondiffell1fixbeta} for the case $\ell=1$, $\beta=1$.
As expected from the familiar case with $V=0$, we find that the large black hole always has a lower action than the small black hole.  Furthermore, in direct analogy with the results in \cite{Marolf:2021kjc}, we find the action of the large wormhole to always be less than that of the small wormhole. In addition, we find Hawking-Page-like transitions for both the black hole and the wormhole branches. Specifically, 
for $\ell=1$ the least-action solution is the large black hole for $V<1.4802(7)$, the thermal AdS solution for  $1.4802(7) <V<3.5785(5)$, and the large wormhole for $V>3.5785(5)$.

The corresponding thresholds for $\ell=2\,,\beta=1$, occur at  $V\approx 2.3588(5)$ and  $V\approx 3.3894(0)$.  In particular, 
We find a similar transitions for both values of $\ell$ despite the fact that black holes exist at arbitrarily large values of $V$ when $\ell=2$.

Having discussed the results for fixed $\beta$, we now explore the behavior of solutions as we vary $\beta$ and $\ell$ within a slice of parameter space with constant $V$.

\begin{figure}[h!]
    \centering
    \includegraphics[width=0.99\linewidth]{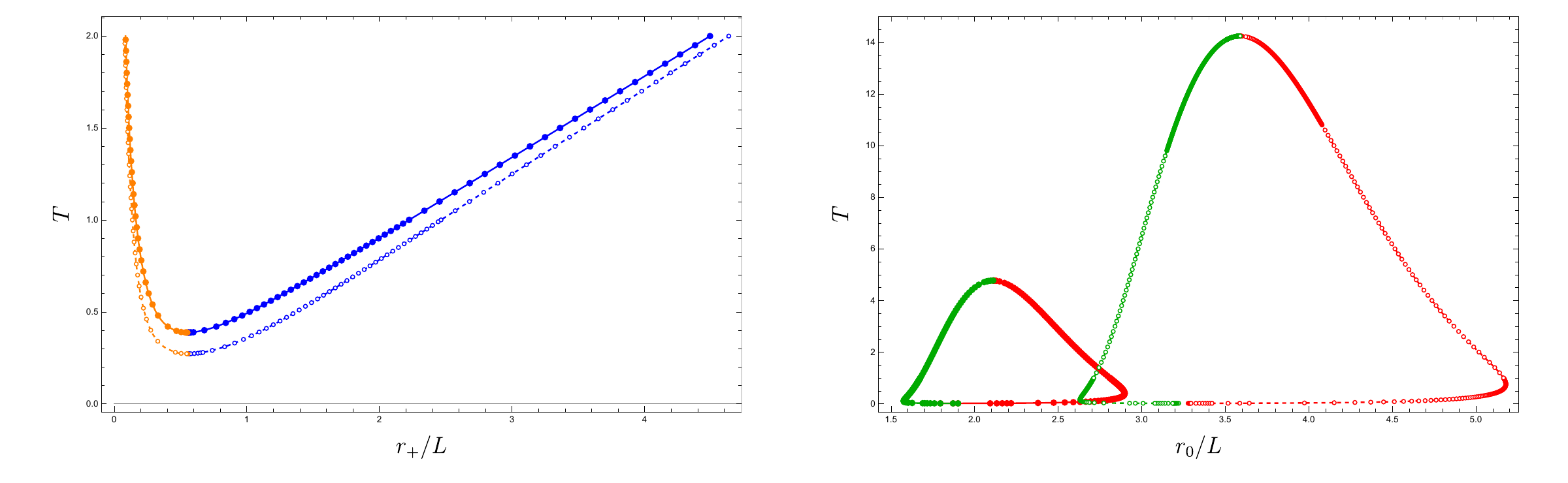}
    \caption{The left and right panels respectively show black hole and wormhole solutions at fixed $V$ for both $\ell=1$ and $\ell=2$.  In the first case we choose $V=1$ so that  black holes exist for both values of $\ell$.  In the second we choose $V=7/2$ so that we find wormholes for both values of $\ell$. The vertical axis is the effective `temperature' defined by $T=\beta^{-1}$.   The color coding is the same as that in figure \ref{fig:fxibeta}.  The solid/dashed lines (solid/open dots) are data for $\ell=1$ and $\ell=2$, respectively.      }
    \label{fig:fixV}
\end{figure}

Results for both $\ell=1$ and $\ell=2$ are shown in figure \ref{fig:fixV} for black hole solutions (with V=1) and wormhole solutions (with $V=3.5$). As for the familiar case $V=0$, black holes exist only above a threshold temperature at which the large and small black hole branches meet. For $V=1$, this threshold temperature is $T=0.3872(4)$ for $\ell=1$ and $T=0.2716(6)$ for $\ell=2$. We also find a Hawking-Page-like transition, meaning that the large black hole dominates over the thermal AdS solution when the temperature is large enough. For $\ell=1$, the least-action solution is the black hole for $T>0.4858(4)$. For $\ell=2$ the transition happens at $T=0.3448(5)$. 

\begin{figure}
    \centering
    \includegraphics[width=0.99\linewidth]{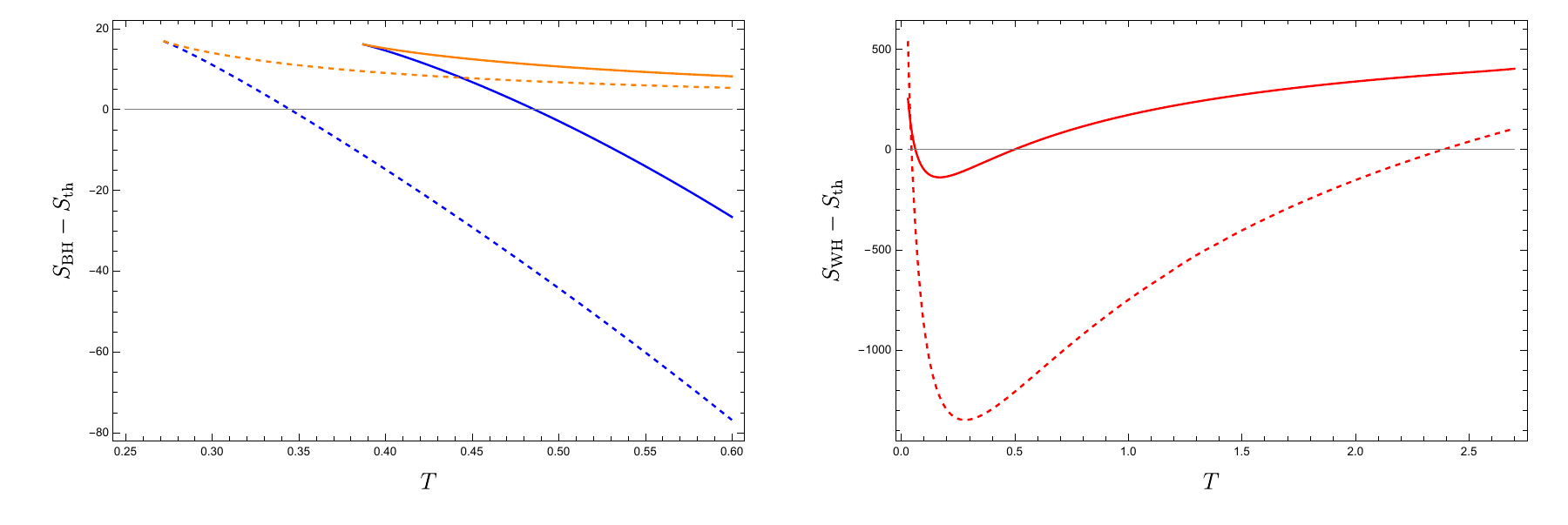}
    \caption{Comparisons of the actions $S_{\mathrm{BH}},S_{\mathrm{WH}}$ and $S_{\mathrm{th}}$ of black holes (left panel) or large wormhole (right panel) vs. the thermal AdS solution for fixed $V$. 
    Phase transitions occur when the colored curves cross the gray lines. As in figure~\ref{fig:fixV}, the orange/blue colors denote small/large black holes, and the red color denotes large wormhole. The solid/dashed lines are for $\ell=1$ and $\ell=2$, respectively.  }
    \label{fig:fixVaction}
\end{figure}

For a fixed $V$ where there are wormhole solutions, they exist only for a finite range of temperatures. For $V=7/2$, we find wormholes exist when $0.01390(0)\le T\le 4.7715(6)$ for $\ell=1$ and $0.01626(0)\le T\le 14.2433(0)$ for $\ell=2$. At the threshold temperatures, the large and small wormhole branches merge. The small wormhole always has a larger action than the large wormhole and never dominates the ensemble. The large wormhole dominates the ensemble when it has a smaller action than the thermal AdS solution. For $\ell=1$, we find that wormholes dominate when $0.0627(1)<T<0.4995(3)$. For $\ell=2$, wormholes have a smaller action than the thermal AdS solutions when $0.0456(0)<T<2.3884(5)$. The action difference between the solutions is shown in figure \ref{fig:fixVaction}.

We are finally ready to present our results regarding the full phase diagram for general $V$ and $T=1/\beta$. These results are shown in figure \ref{fig:phase} below. The solid and dashed lines mark the boundaries of the regions in which we find black hole/wormhole solutions. The shaded regions show where each solution has the least action and is thus dominate the ensemble. In particular, for $\ell=1$, black holes minimize the action in the green shaded region, while wormholes have minimal action in the purple shaded region. In the region between the green and purple shaded regions, the thermal AdS solution has an action that is less than that of either black holes or wormholes.  The phase structure is similar for $\ell=2$, black holes/wormholes dominate the ensemble in the orange/blue shaded regions (which we take to also include the green/purple shaded regions), and thermal AdS dominates in the unshaded region.

\begin{figure}
    \centering
    \includegraphics[width=0.6\linewidth]{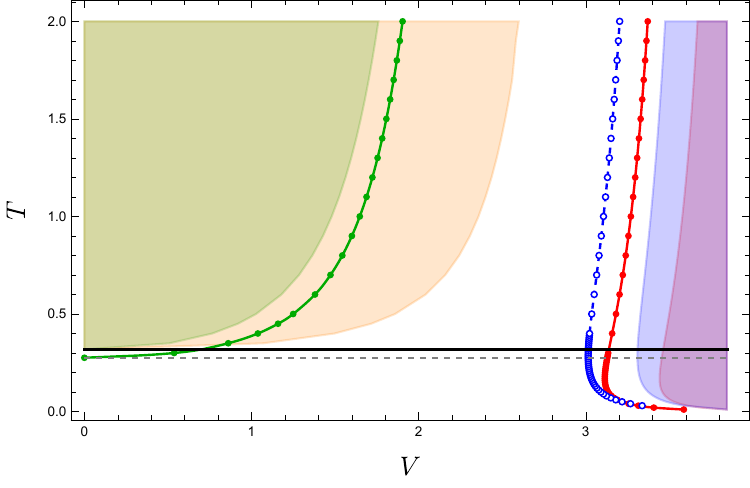}
    \caption{A phase diagram for our Einstein-scalar model. The horizontal dashed gray line marks the minimum temperature $T_c=\sqrt{3}/(2\pi)$, above which we find black hole solutions, while the horizontal black line  marks the temperature $T_{\rm HP}=1/\pi$ of the $V=0$
    Hawking-Page transition. The green dots show the maximum $V$ at each $T$ for which we find black hole solutions when $\ell=1$. In contrast, for $\ell=2$ and $T\ge T_c$ we find black holes for all values of $V$. The red/blue dots mark the minimum $V$ required to support a wormhole for $\ell=1$ and $\ell=2$, respectively. For $\ell=1$, black hole/wormhole solutions dominate in the green/purple shaded region, and thermal AdS solutions dominate in the region between them. For $\ell=2$, in addition to the regions in which they dominated for $\ell=1$, black hole/wormhole solutions also dominate in the orange/blue regions. Thermal AdS solutions dominate in the unshaded region in between. 
   }
    \label{fig:phase}
\end{figure}

\section{Adding a source with $\ell=0$}\label{sec:ell0}

We now turn to the question of whether wormholes are generic at large source amplitude and fixed mass.  As we have seen, wormholes in fact dominate at large $V$ for $\ell \ge 1$.  However, as described in the introduction, they are forbidden by topological censorship for $\ell=0$. We emphasize that this is the case for any value of the momentum $n$ on the $S^1$.   If such results are to be ascribed to some non-generic behavior, then in the limit of a large $\ell=0$ source (with fixed mass), wormholes would need to reappear (and, in fact, to dominate) as soon as we turn on a small $\ell\ge 1$ perturbation. 

To explore this possibility, we now add an additional complex scalar field $\Phi_0$ to the model studied in section \ref{sec:EGCS}.  We also rename $\vec \Phi$ to $\vec \Phi_1$.   We then take $\Phi_0$ to have angular momentum $n=1$ on the $S^1$ but to have $\ell=0$ on the $S^2$,  while we take $\vec \Phi_1$ to have both $n=1$ and $\ell=1$, so that the value of $\ell$ matches the label on the field. The amplitude of the source for each field will be denoted $V_0,V_1$.

The action then takes the form
\begin{equation}
\label{eq:action2}
S=-\frac{1}{16\pi G}\int_{\mathcal{M}}\mathrm{d}^4x\sqrt{g}\left(R+\frac{6}{L^2}-2\nabla^a \vec{\Phi}_1 \cdot \nabla_a\vec{\Phi}_1^\star-2\nabla^a {\Phi}_0  \nabla_a{\Phi}_0^\star\right)-\frac{1}{8\pi G}\int_{\partial \mathcal{M}}\mathrm{d}^3x\sqrt{h}K-S_{\partial{\mathcal{M}}}\, ,
\end{equation}
where ${}^\star$ denotes complex conjugation, $\cdot$ denotes the usual Cartesian dot product between vectors in $\mathbb{R}^{2\ell+1}$, $L$ is the AdS length scale. 
The equations of motion are
\begin{subequations}
\begin{equation}
\begin{split}
    R_{ab}-\frac{R}{2}g_{ab}-\frac{3}{L^2}g_{ab}=&\nabla_a \vec{\Phi}_1\cdot\nabla_b \vec{\Phi}
    _1^\star+\nabla_b \vec{\Phi}_1\cdot\nabla_a \vec{\Phi}_1^\star-g_{ab}\nabla_c \vec{\Phi}_1\cdot\nabla^c \vec{\Phi}_1^\star\\
    +&\nabla_a {\Phi}_0\nabla_b {\Phi}_0^\star+\nabla_b {\Phi}_0\nabla_a {\Phi}_0^\star-g_{ab}\nabla_c {\Phi}_0\nabla^c {\Phi}_0^\star\,,
\end{split}
\end{equation}
and
\begin{equation}
\nabla_a \nabla^a \vec{\Phi}_1=0\,, \quad \nabla_a\nabla^a \Phi_0=0\,.
\end{equation}
\end{subequations}
Note that since spherical harmonics with different $\ell$ are orthogonal on the sphere, the action (and thus the equations of motion and solutions) would be identical if we simply gave one of the original fields in $\vec \Phi_1$ an $\ell=0$ component with amplitude $\Phi_0$.

We will again consider boundary metrics $g^{[0]}_{\mu\nu}$ of the form 
\begin{equation}\label{eq:bdymet}
\mathrm{d}s^2_{\partial \mathcal{M}}=\mathrm{d}\tau^2+L^2 \mathrm{d}\Omega_2^2
\end{equation}
where
\begin{equation}
\mathrm{d}\Omega_2^2=\mathrm{d}\theta^2+\sin^2\theta\mathrm{d}\varphi^2
\end{equation}
is the standard round metric on a unit radius two-sphere, with $\theta\in[0,\pi]$, $\varphi\sim\varphi+2\pi$ and $\tau\sim\tau+\beta\,L$.
Our solutions thus depend on $(\ell,V_1,V_0,\beta)$.

We will also consider adding a component with $\ell=2$ instead of $\ell=1$, denoting the associated source amplitude by $V_2$. Higher angular momenta can  also be treated analogously.

We can find wormhole solutions using a procedure similar to that described in appendix \ref{app:conwh} for the case without the $\ell=0$ term. As described in section \ref{sec:pwellzero}, their stability properties are analogous to those of the wormholes without the $\ell=0$ term.  

\begin{figure}
    \centering
    \includegraphics[width=0.99\linewidth]{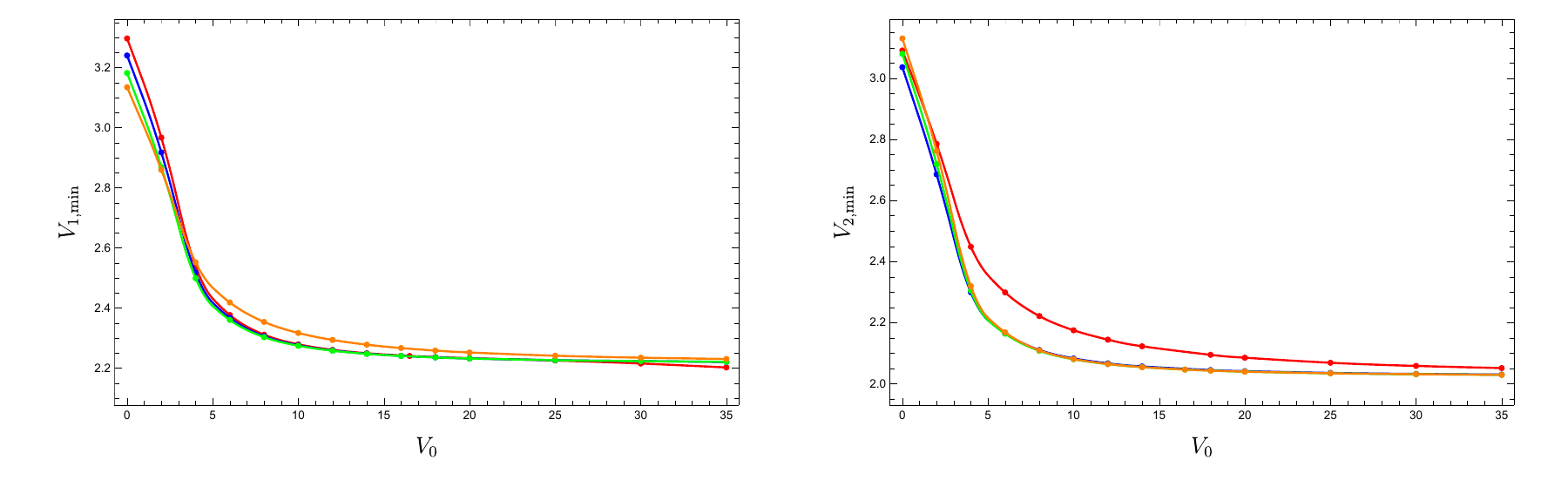}
    \caption{The left/right panel  shows the minimum $V_1/V_2$ required to support a wormhole at fixed temperature $T$ for various $V_0$ for $\ell=1$/$\ell=2$, respectively. The four temperatures here are $T=6/5$ (shown in red), $T=4/5$ (shown in blue), $T=1/2$ (shown in green), and $T=1/10$ (shown in orange).}
    \label{fig:fixT}
\end{figure}

While we wish to study wormholes of fixed mass, it is the Euclidean period over which we have explicit control. We therefore first generate a large collection such wormholes at various temperatures and then sort them into subsets with the same value of the mass.  The data for fixed temperatures is displayed in figure \ref{fig:fixT} for cases where the symmetry-breaking perturbation has $\ell=1$ and $\ell=2$.   Note that for $\ell=1$ the red curve (highest temperature) 
eventually enjoys the smallest value of $V_{1,\rm{min}}$ at high enough $V_0$.   Similarly, the orange curve (smallest temperature)  
eventually enjoys the largest value of $V_{1,\rm{min}}$ at high enough $V_0$. However, neither of these are the case 
until near the right edge of out plot (where the wormholes become more difficult to construct).  
Since the problem of interest involves the limit of large $V_0$, we expect that it is important to probe large enough values of $V_0$ such that this transition has occurred.   Examining the right plot in figure \ref{fig:fixT}, it seems likely that a similar transition will occur for $\ell=2$, but that we were not able to probe sufficiently high values of $V_0$ due to limited computing resources.  We therefore consider only the case $\ell=1$ in studying fixed-mass below, though it seems likely that the large $V_0$ limit is similar for $\ell \ge 2$.

Turning now to the fixed-mass case, we must first find an expression for the desired mass function.  This is done in appendix \ref{app:wmass} by writing
the compact coordinate $y$ described there as a Taylor series in the Fefferman-Graham coordinate $z$ near conformal infinity $(z=0)$.  This procedure allows us to extract the mass of our wormholes from the FG expansion.  The result is given in \eqref{eq:massEq}.

\begin{figure}
    \centering
    \includegraphics[width=\linewidth]{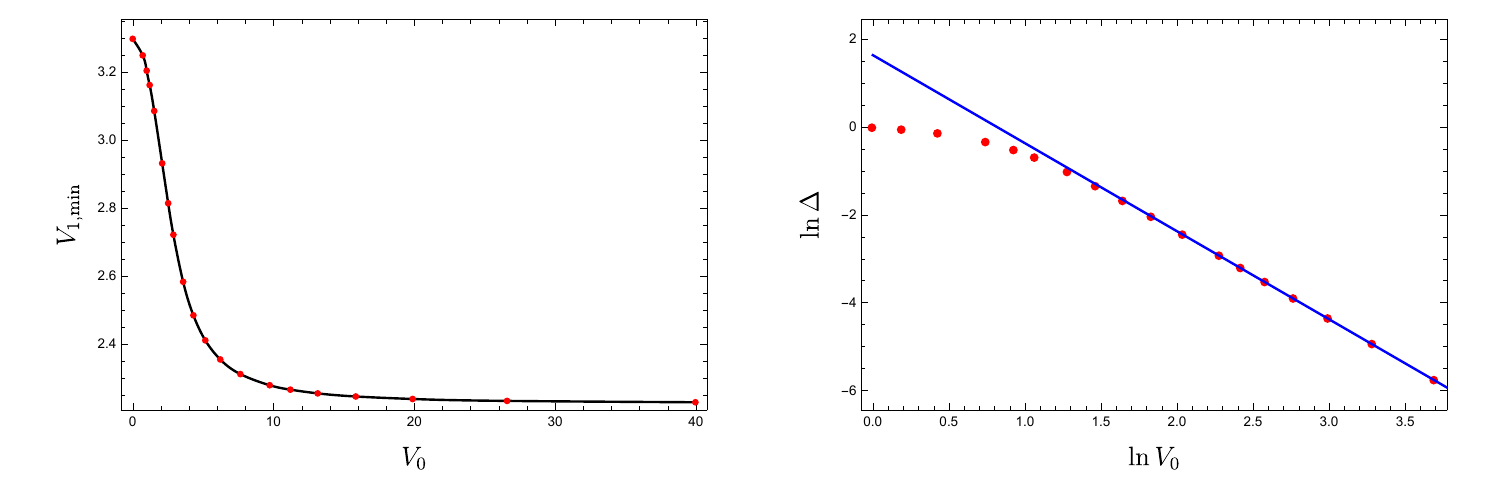}
    \caption{The minimum $V_1$ required to support a wormhole with fixed mass for various $V_0$ is shown at left.  The chosen mass is that of the critical wormhole with $T=6/5$ and $V_0=0$. The right panel shows  a log-log plot of the data points (red) and the 3-parameter best-fit power law \eqref{eq:bestfitfixedmass} (blue line), where $\Delta\equiv V_{1,\mathrm{min}}-2.22458$.  The results suggest that $V_{1,\mathrm{min}}$ fails to vanish in the limit $V_0\to\infty$.}
    \label{fig:V1minVSV0}
\end{figure}

Our main results are shown in figure \ref{fig:V1minVSV0}. Recalling that we wish to fix the total mass, we chose a mass corresponding to that of the critical wormhole at $T=6/5$ with $V_0=0$. By critical wormhole, we mean the solution at which the large and small wormhole branches merge. The minimum $V_1$ required to support a wormhole of this mass for various $V_0$ is displayed in the left panel. While we cannot numerically study the case where $V_0$ is strictly infinite, we performed a 3-parameter power law fit to data shown, and using the last $5$ data points, we find 
\begin{equation}
\label{eq:bestfitfixedmass}
    V_{1,\mathrm{min}}=2.22458+5.09883 \,V_0^{-2.00378}\,.
\end{equation}
As shown in the right panel, this is indeed an excellent fit to our data.  We thus find strong evidence that the minimum $V_1$ required to support a wormhole remains non-zero in the $V_0\to\infty$ limit, and thus that wormholes are {\it not} in fact generic at large $V_0$ in our cohomogeneity-1 model.

\section{Perturbative stability of wormholes}
\label{sec:stability}

We now consider the perturbative stability of the above wormhole solutions in the sense of  investigating whether the quadratic action around each solution has negative modes. We allow general perturbations with arbitrary momenta on both the $S^1$ and the $S^2$.

In direct analogy with the results of \cite{Marolf:2021kjc}, we find that the large wormhole has no negative mode, and is thus stable, while the small wormhole is unstable due to the existence of a single negative mode.

Any attempt to address stability must face the fact that the Euclidean gravitational action can be arbitrarily negative, even with fixed boundary conditions \cite{Gibbons:1978ac}.  As a result, the integral over all real Euclidean metrics fails to converge and thus does not define a meaningful theory.  It is typically assumed that one must therefore integrate over some contour through the space of complex deformations of Euclidean-signature metrics, though the prescription for choosing the `correct' contour is currently far from fully understood\footnote{However, we are sympathetic to the point of view that the defining contour should be taken to be given by real Lorentzian metrics   \cite{Hartle:2020glw,Schleich:1987fm,Mazur:1989by,Giddings:1989ny,Giddings:1990yj,Marolf:1996gb,Dasgupta:2001ue,Ambjorn:2002gr,Feldbrugge:2017kzv,Feldbrugge:2017fcc,Feldbrugge:2017mbc}, and in particular in the form described in \cite{Marolf:2020rpm,Colin-Ellerin:2020mva,Colin-Ellerin:2021jev,Marolf:2022ybi}.}.
In the absence of such understanding, we simply choose to use a common method of analyzing the stability of saddles that follows \cite{Garriga:1997wz,Gratton:1999ya,Kol:2006ga} in choosing a complex contour adapted to some foliation of the spacetime.  In particular, recalling that that any such foliation is associated with gravitational constraints, one chooses a contour such that the integral over the associated lapse and shift converges.  In the canonical version of this procedure (proposed in 
\cite{Garriga:1997wz,Gratton:1999ya}), one then obtains a set of delta-functions that impose the constraints.  However, in the covariant version of this procedure (introduced in \cite{Kol:2006ga}), one performs the lapse and shift integrals in the stationary phase approximation, effectively solving the constraints for lapse and shift.  

This, however, leads to an interesting subtlety when applied to ${\mathbb Z}_2$-symmetric wormholes using the natural radial foliation.   Since the Hamiltonian constraint is quadratic in extrinsic curvatures, the kinetic term in the linearized constraint vanishes identically at such a surface.  As a result, one cannot solve the constraint to express the associated radial lapse in terms of other fields and derivatives. We will see that this leads to the apparent appearance of a divergence in the action at the  ${\mathbb Z}_2$-invariant surface.  However, we will also see that this apparent divergence is removed when one realizes that, also due to the above degeneracy, integrating over the lapse on this particular surface leads to a delta-function that imposes the linearized Hamiltonian constraint.  This procedure was implicitly used previously without comment in \cite{Marolf:2021kjc}. Related divergences also appear in the canonical version of the procedure (following \cite{Garriga:1997wz,Gratton:1999ya}) and have plagued various studies of wormhole stability though, as shown recently in \cite{Marolf:2025evo}, they again cancel when treated with similar care.

Before proceeding, we pause to emphasize again that there is no known fundamental derivation of the above procedure. 
 In addition, specific concerns about this approach were recently raised in \cite{Horowitz:2025zpx}.  For these reasons, in appendix \ref{app:ROT} we perform a second study of stability using an alternative prescription from \cite{Marolf:2022ntb} that generalizes the Wick-rotate-the-trace-mode prescription of \cite{Gibbons:1978ac}. Happily, the results of both methods agree.

Although we work in Euclidean signature, as noted above
it is useful to write the metric in the language of $(3+1)$-decomposition, taking the radial direction to play the role of Euclidean time.   
The perturbed metric can be written in the form 
\begin{equation}
    \dd s^2=g_{\tau\tau}(y,\tau)\dd\tau^2+2\beta(y,\tau)\dd \tau\dd y+\left[\alpha^2(y,\tau)+\frac{\beta^2(y,\tau)}{g_{\tau\tau}(y,\tau)}\right]\dd y^2+g_{\theta\theta}\dd\Omega_2^2\,,
\end{equation}
where $\alpha$ and $\beta$ are radial-versions lapse and shift and $y\in[0,1)$ is a compact coordinate defined by 
\begin{equation}
    y=\frac{r}{r+r_0}\Leftrightarrow r = \frac{yr_0}{1-y}\,.
\end{equation}

Before proceeding, let us recall
that our wormhole solutions preserve an $\mathrm{U}(1) \times \mathrm{SO}(3)$ symmetry that acts by rotating of the $S^1\times S^2$ factor in the metric and by simultaneously acting on the space of scalars.  Organizing perturbations into representations of this symmetry then gives modes that decouple from one another. The simplest such sector is the one that preserves the full $\mathrm{U}(1) \times \mathrm{SO}(3)$ symmetry.   We discuss this sector explicitly in detail in section \ref{sec:poaformalism}.  Since the other sectors are conceptually similar but technically more complicated, their treatment is relegated to appendix \ref{app:sectors}.

\subsection{The static spherical sector}
\label{sec:poaformalism}

In the sector preserving our full diagonal $\mathrm{U}(1) \times \mathrm{SO}(3)$ symmetry,  the metric perturbation has no time dependence, and the scalar perturbation has a time dependence that is the same as that of the background scalar field. For this case, we use the following  ansatz for the metric perturbation,
\begin{equation}\label{eq:pertm0g}
    \delta\dd s^2= \frac{\epsilon}{(1-y)^2}\left[\frac{r_0^2}{L^2}(1+2y^2-2y)\delta q_1(y)\mathrm{d}\tau^2+\frac{L^2 \cdot \delta q_2(y)}{1+2y^2-2y}\mathrm{d}y^2+r_0^2(1+2y^2-2y)\delta q_4(y)\mathrm{d}\Omega_2^2\right]\,,
\end{equation}
while taking the scalar field perturbation to be of the form
\begin{equation}\label{eq:pertm0phi}
    (\delta \vec{\Phi})_m=\delta q_3(y)e^{-i \frac{2\pi n}{\beta\,L} \tau}\times \sqrt{\frac{(\ell-m)!}{(\ell+m)!}}\times P^{\ell}_m(\cos \theta)\times\left\{
\begin{array}{ll}
\sqrt{2}\,\sin (m\,\varphi)\,,& \text{for}\quad m<0
\\
1\,, & \text{for}\quad m=0
\\
\sqrt{2}\,\cos (m\,\varphi)\,,& \text{for}\quad m>0
\end{array}\right.\,,
\end{equation}
At the conformal infinity $y=1$, we impose asymptotic AdS boundary conditions.  These conditions require
\begin{equation}
    \delta q_i(1)=\delta q_i'(1)=0\,,\quad i=1,2,3,4\,.
\end{equation}

Instead of studying the full wormhole, it is convenient to make use of the $\mathbb {\mathbb Z}_2$ reflection symmetry about the wormhole throat.  We therefore decompose the space of perturbations into even and odd sectors under this $\mathbb {\mathbb Z}_2$ and study only the behavior on half of the wormhole.
At the wormhole neck we then impose 
\begin{equation}
    \delta q_i(0)=0\quad \mathrm{in\ the\ odd\ sector}\,,\quad\mathrm{or}\quad \delta q_i'(0)=0\quad \mathrm{in\ the\ even\ sector}\,.
\end{equation}

With this ansatz for the perturbations, we evaluate the action~\eqref{eq:action1} to quadratic order in $\epsilon$ to construct a quadratic action $S^{[2]}$. 
 This $S^{[2]}$ is a function of the $\delta q_i\,,\ i=1,2,3,4$ and their first and second derivatives with respect to $y$.  Integrating the 2nd derivatives by parts yields a boundary term at the conformal boundary which,  after imposing boundary conditions, precisely cancels the variation of the explicit boundary terms in our action. What remains is an action that contains only $\delta q_i$ and the first $y$-derivatives $\delta q_i'$\,. One can then perform further integrations by parts to remove all remaining derivatives from every appearance of $\delta q_2$; i.e., so that $\delta q_2$ then appears algebraically.  This is a consequence of the Bianchi identities, and it is associated with the fact that $\delta q_2$ is the linearized `radial lapse' and since for each mode depends only on our radial coordinate $y$.

 We denote the resulting action by $S_A^{[2]}$\, where the subscript indicates the algebraic appearance of $\delta q_2$.   This action takes the form
\begin{equation}
    S_A^{[2]}=\int_0^1 \dd y \left(A\cdot \delta q_2^2+B\cdot \delta q_2+C \right),
\end{equation}
where $A$ depends only on background fields, $B$ is linear in the perturbations $\delta q_1,\delta q_3,\delta q_4$ and their first derivatives (but independent of $\delta q_2$), and $C$ is quadratic in $\delta q_1,\delta q_3,\delta q_4$ and their first derivatives (but independent of $\delta q_2$).  In particular, the function $A$ arises from the extrinsic curvature squared terms in the radial Hamiltonian constraint, as other terms in the constraint appear in the action in a form that is manifestly linear in the lapse.  As a result, $A=0$ on constant-$y$ hypersurfaces with vanishing extrinsic curvature\footnote{Since the kinetic term is not positive definite, it is in general possible for $A$ to vanish on other hypersurfaces as well.  However, this does not occur in our solutions.}.  In our context, this occurs only at the wormhole neck, where it is required by our imposition of ${\mathbb Z}_2$ symmetry. 

As mentioned above, the prescription of \cite{Kol:2006ga} chooses a contour that allows us to  perform the Gaussian integral over $\delta q_2$. 
For $A\neq 0$, the stationary point corresponds to solving the linearization of the radial Hamiltonian constraint for $\delta q_2$.  We may thus formally write the result in terms of an action for the remaining variables that the form 
 \begin{equation}
\label{eq:solvedq2}     S_C^{[2]}=-\int_0^1 \dd y \left[\frac{B^2}{4A}+C\right]\, ,
 \end{equation}
 where the subscript indicates that we have now imposed the linearized constraint.

Now, as previously advertised, the reader will note that the right-hand-side of \eqref{eq:solvedq2} contains an explicit divergence at the wormhole neck where $A=0$. However, we should realize that, since $A=0$ at the wormhole neck $(y=0)$, the action $S_C^{[2]}$ was {\it not} in fact quadratic there.  As a result, the contour that makes integration over $\delta q_2(0)$ well-defined involves integrating $\delta q_2(0)$ over the imaginary axis.  Performing that integral thus yields a delta function that requires $B(0)=0$.  Treating this carefully as a new boundary condition on $\delta q_1,\delta q_3, \delta q_4$ removes the divergence and yields a finite action.   This procedure was also used without explicit comment in \cite{Marolf:2021kjc}, though see \cite{Marolf:2025evo} for discussion of how related apparent divergences have plagued studied of wormhole stability that follow a canonical version of the above procedure and how they can again be shown to cancel.

Our action will necessarily be invariant under all linearized diffeomorphisms that preserve our $\mathrm{U}(1)\times \mathrm{SO}(3)$ symmetry.  The most general such diffeomorphism is generated by a vector field which, when we lower an index, defines a one-form $\xi=L^2\xi_y(y)\dd y$.
The associated gauge transformations take the form:
\begin{subequations}
    \begin{equation}
        \delta q_1(y)=(1-y)q_2(y)\left\{2yq_1(y)+(1-y)[1-2(1-y)y]q_1'(y)\right\}\xi_y(y)\,,
    \end{equation}
    \begin{multline}
        \delta q_2(y)=-(1-y)\left[4-10y+8y^2+(2 y^3-4 y^2+3 y-1)\frac{q_2'(y)}{q_2(y)}\right]\xi_y(y)
        \\
        +2(1-y)^2(1+2y^2-2y)\xi_y'(y)\,,
    \end{multline}
    \begin{equation}
        \delta q_4(y)=2(1-y)y q_2(y)\xi_y(y)
    \end{equation}
    \begin{equation}
        \delta q_3(y)=(1-y)^2(1-2y+2y^2)q_2(y)q_3'(y)\xi_y(y)\,,
    \end{equation}
\end{subequations}
where $'$ denotes derivatives with respect to $y$.

We expect that we can find gauge-invariant linear combinations of our perturbative fields $\delta q_i$ that are invariant under the above transformations, and also that our action $S^{[2]}_C$ can be understood as a function of the resulting gauge-invariant variables.  Since we have three remaining $\delta q$'s and there is one gauge symmetry, we find two gauge invariant variables:
\begin{subequations}
    \begin{equation}
    \label{eq:P1}
        P=\delta q_3(y)+\frac{(y-1) [2 (y-1) y+1] q_3'(y)}{2 y}\delta q_4(y)\,,
    \end{equation}
    \begin{equation}
    \label{eq:Q1}
        Q=\delta q_1(y)-\left[q_1(y)+\frac{(1-y)(1-2y+2y^2)q_1'(y)}{2y}\right]\delta q_4(y)\,,
    \end{equation}
\end{subequations}
which satisfy boundary conditions
\begin{equation}
    \begin{split}
        &P(1)=P'(1)=Q(1)=Q'(1)=0\,,
    \end{split}
\end{equation}
and 
\begin{equation}
    \begin{split}
       & P(0)=Q(0)=0\quad \mathrm{in\ the\ odd\ sector}\,,\\\quad\mathrm{or}\quad & P'(0)=Q'(0)=0\quad \mathrm{in\ the\ even\ sector}\,.
    \end{split}
\end{equation}
Solving \eqref{eq:P1} and \eqref{eq:Q1} for $\delta q_3$ and $\delta q_1$ and substituting the results into $S_C^{[2]}$ gives an expression in involving only $P,Q,\delta q_4$ and their derivatives. After further integrations by parts and repeated use of the boundary conditions, the dependence of the various terms on $\delta q_4$ can then be shown to cancel as expected.  The final result then writes $S_C^{[2]}$ as a  expression in terms of $P,Q$,  their derivatives, and the background fields.  Since the expression is rather complicated, we refrain from displaying it explicitly.

A standard method of analyzing positivity of $S_C^{[2]}$ is turn turn the second derivatives into a self-adjoint operator $\mathcal{L}$ and to then diagonalize $\mathcal{L}$.  But  such an $\mathcal{L}$ is a $(1,1)$ tensor on the space of perturbations, while the second derivatives of $S_C^{[2]}$ naturally define a $(2,0)$ tensor.  We thus need to choose an inner product that can be used to raise one of the indices.  If the inner product is positive-definite, then it also defines a Hilbert space in which the resulting $\mathcal{L}$ is automatically symmetric (see e.g. \cite{Marolf:2022ntb}) and, since it takes a standard Sturm-Liouville form, it is in fact self-adjoint.    We will choose the simple inner product
\begin{equation}
\label{eq:IP1}
    (\vec Q_1,\vec Q_2)=\int_{\mathcal{M}}\dd^4 x\sqrt{g}\, \vec Q_1\cdot \vec Q_2\ = \int_{\mathcal{M}}\dd^4 x\sqrt{g}\, \left(P_1 P_2+ Q_1Q_2\right)\,
\end{equation}
on the space of gauge-invariant perturbations
$\vec Q=(P,Q)$\,. 

\begin{figure}[h!]
    \centering
    \includegraphics[width=0.9\linewidth]{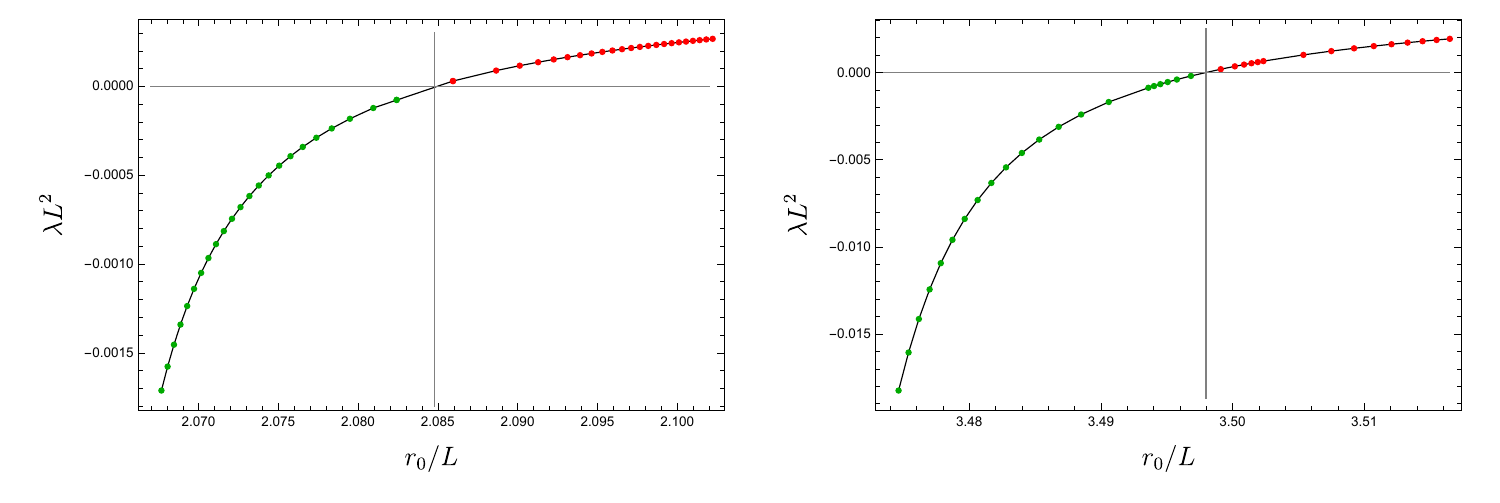}
    \caption{The lowest lying mode of $\mathcal{L}$ in the even sector.  The left/right panels show $\ell=1$ and $\ell=2$ with $n=\beta=1$. The horizontal lines are $\lambda=0$ while the vertical lines mark the critical value of $r_0$ from figure~\ref{fig:fxibeta}. The $\lambda<0$ green dots to the left are thus small wormholes while the $\lambda >0$ red dots to the right are large wormholes.}
    \label{fig:spec_gaugeinv}
\end{figure}

We find a single negative mode for the small wormholes in the even $m=0$ sector. The lowest lying mode for $\mathcal{L}$ in this sector is shown in figure \ref{fig:spec_gaugeinv}\,.

The same method can also be applied to analyze stability in the other sectors.  The details are presented in 
Appendix~\ref{app:sectors}.
In particular, Appendix \ref{app:sectors} gives the explicit perturbation ansatz and the construction of gauge-invariant variables for each sector. For some sectors it is possible to show analytically (or with minimal use of numerics) that there are no negative modes for either the small or large black hole.   For the remaining sectors, Appendix \ref{app:sectors} presents numerical evidence that there are no further negative modes.   In particular,  the dependence of the lowest eigenvalue on the neck radius for $\ell=1$ wormhole 
backgrounds in the scalar-derived sectors is shown in Figure~\ref{fig:higherspec} of Appendix~\ref{app:D7}.

\subsection{Adding an $\ell=0$ source}
\label{sec:pwellzero}

\begin{figure}
    \centering
    \includegraphics[width=0.98\linewidth]{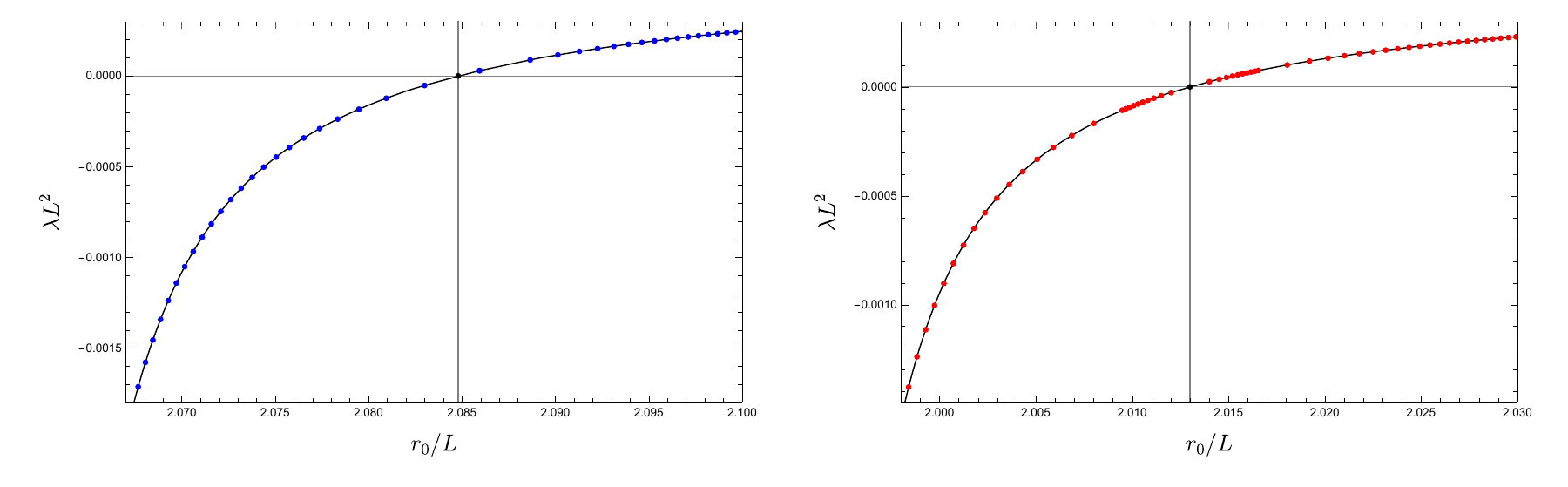}
    \caption{The lowest eigenvalue of ${\cal L}$ as a function of the wormhole neck radius for backgrounds with $n=\beta=1$. The left/right panel has $V_0=0$ and $V_0=1$, respectively. The horizontal gray line denotes $\lambda=0$, and the vertical gray line denotes the value of $r_0$ where the large and small wormholes merge as $V_1$ tends to the critical minimal value. The black dots denotes the position where the radius is the same as the critical radius, and eigenvalue is zero. }
    \label{fig:TwoScalarSpec}
\end{figure}

We can also study stability for these wormholes using the method described above. Again we find the small wormhole is unstable and has a negative mode, while the large wormhole is stable and has no negative modes. The results for $V_0=0$ and $V_0=1$ are shown in figure \ref{fig:TwoScalarSpec}. Here we have checked only perturbations of the form
\begin{equation}
    \begin{split}
         \delta\dd s^2&= \frac{\epsilon}{(1-y)^2}\left[\frac{r_0^2}{L^2}(1+2y^2-2y)\delta q_1(y)\mathrm{d}\tau^2+\frac{L^2 \cdot \delta q_2(y)}{1+2y^2-2y}\mathrm{d}y^2+r_0^2(1+2y^2-2y)\delta q_4(y)\mathrm{d}\Omega_2^2\right]\,,\\
         (\delta \vec{\Phi}_1)_m&=\delta q_{3,1}(y)e^{-i \frac{2\pi n}{\beta\,L} \tau}\times \sqrt{\frac{(1-m)!}{(1+m)!}}\times P^{1}_m(\cos \theta)\times\left\{
    \begin{array}{ll}
    \sqrt{2}\,\sin (m\,\varphi)\,,& \text{for}\quad m=-1
    \\
    1\,, & \text{for}\quad m=0
    \\
    \sqrt{2}\,\cos (m\,\varphi)\,,& \text{for}\quad m=1
    \end{array}\right.\,,\\
    \delta \vec{\Phi}_0&=\delta q_{3,0}(y)e^{-i \frac{2\pi n}{\beta\,L} \tau}\,,
    \end{split}
\end{equation}
similar to the case considered in section \ref{sec:poaformalism}. Note that the case $V_0=0$ is the same as the $\ell=1$ case in figure \ref{fig:spec_gaugeinv}.  

\section{Discussion}
\label{sec:concl}

Our main task was to begin to probe a conjecture of \cite{Balasubramanian:2022gmo}, namely that inserting distinct operators of the form $e^{-\beta H} A_\beta$ on each half of the Euclidean boundary will generically lead to wormholes dominating in a limit $\beta \rightarrow \infty$ in which the energies of the two resulting states are held fixed.  However, as reviewed in the introduction, for wormholes supported by appropriate field-theoretic sources, one can show that there can be no wormholes unless all $\mathrm{U}(1)$ symmetries are broken.  This includes $\mathrm{U}(1)$ symmetries that are part of a larger rotation group $\mathrm{SO}(N)$ for $N>2$.  Since there are many contexts in which one can introduce large sources that preserve such symmetries, the above conjecture requires that doing so leads either to spontaneous symmetry breaking in the bulk, or to the nucleation of wormholes under the addition of arbitrarily small symmetry-breaking sources.

We studied a version of this conjecture for Einstein-Hilbert gravity coupled to scalar fields in a particular ansatz with $S^1 \times S^2$ boundaries and for which bulk solutions are cohomogeneity-1.   The role of the operator $A_\beta$ is played by the boundary condition for our scalars.
Interestingly, we find no evidence that wormholes generically dominate in the desired limit.  In particular,
if we restrict the sources to be built only from angular momenta $\ell =0$ and $\ell=1$ on the $S^2$, we find that wormholes exist only when the part of the source with $\ell=1$ exceeds a critical strength $V_1$ that is largely independent of the strength $V_0$ of the $\ell=0$ part of the source; see again figure \ref{fig:V1minVSV0}.   We also find similar results when the $\ell=1$ component is replaced by an $\ell=2$ component, and we again expect similar results for higher angular momenta as well.

Along the way, we mapped out the phase diagram for our cohomogeneity-1 ansatz and studied stability of the wormhole saddles.  Results for the wormholes were largely analogous to those found in
\cite{Garcia-Garcia:2020ttf,Cotler:2020lxj,Marolf:2021kjc}, where wormholes appear only at finite at some source-strength at which the wormhole neck has a certain critical size. At larger sources there are then two wormhole solutions, one with a neck smaller than the critical size and one with a neck larger than the critical size.  As in \cite{Marolf:2021kjc}, we find a single negative mode for the small wormholes and no negative modes for the large wormholes. In particular, our study of wormhole and their perturbations finds no evidence of spontaneous symmetry breaking. The overall structure is thus directly analogous to that of the Hawking-Page transition \cite{Hawking:1982dh}.  In contrast, black hole solutions are suppressed by our scalar sources, though with details that depend strongly on the angular momentum $\ell$ of the source; see again figure \ref{fig:phase}.  

In the main text we studied stability using a method that follows \cite{Garriga:1997wz,Gratton:1999ya,Kol:2006ga}.
As a side note, we then checked in appendix \ref{app:ROT} that the rule-of-thumb prescription for studying stability of Euclidean saddles of \cite{Marolf:2022ntb} (using the DeWitt$_{-1}$ metric) equivalent results.
In the absence of a fundamental derivation of the correct choice of contour for the Euclidean path integral, this serves as a partial check on both methods.  See also \cite{Horowitz:2025zpx} for recent comments on the importance of this issue, and in particular for comments on the method of \cite{Garriga:1997wz,Gratton:1999ya,Kol:2006ga}.

Returning to the issue of genericity of wormholes at large sources, it is important to emphasize that  our codimension-1 ansatz forbids sources from turning off on the $t=0$ slice.  As a result, changing our source strength should also be understood as changing the {\it theory} in which our states live.  Since we tune the sources at each $\beta$ to fix the mass of the solutions, this means that our theory depends on $\beta$ as well. For example, in an AdS/CFT context, we would have chosen certain couplings to depend non-trivially on $\beta$.  This differs substantially from the context described in \cite{Balasubramanian:2022gmo}, where the goal was to take $\beta$ (and $A_\beta$) large within a given theory.  

A strict investigation of their conjecture thus requires a study of wormholes that are cohomogeneity-2 or higher.  Our cohomogeneity-1 study here was largely a warm-up for that more complicated problem, though our negative results emphasize the importance of carrying it out in full.
We therefore hope to address it in the near future.

Finally, we note that an absence of on-shell wormholes for generic large sources need not necessarily mean that wormhole effects are unimportant in that context.  For example, despite being dual to a matrix integral, at fixed $\beta$ pure Jackiw-Teitelboim gravity is known to have no on-shell Euclidean wormholes at all \cite{Saad:2019lba}.  Instead, in JT gravity, wormholes provide important off-shell contributions and, in particular, important endpoint contributions to the gravitational path integral.  We hope to address this possibility (say, for the current cohomogeneity-1 model) in future work as well.

\section*{Acknowledgements} 
We thank Tom~Hartman, Maciej~Kolanowski, Jake~McNamara, Rob~Myers, and Zhencheng~Wang for helpful discussions.   X.~L. and D.~M. were supported in part by NSF grant PHY-2408110 and by funds from the University of California. Use was made of computational facilities purchased with funds from the National Science Foundation (CNS-1725797) and administered by the Center for Scientific Computing (CSC). The CSC is supported by the California NanoSystems Institute and the Materials Research Science and Engineering Center (MRSEC; NSF DMR 2308708) at UC Santa Barbara. Research at Perimeter Institute is supported in part by the Government of Canada through the Department of Innovation, Science and Economic Development and by the Province of Ontario through the Ministry of Colleges and Universities. J.~E.~S. has been partially supported by STFC consolidated grant ST/X000664/1 and by Hughes Hall College.

\appendix

\section{Gauge choices and boundary condtitions}\label{App:construction}
This appendix presents technical details regarding gauge choices and the implementation of boundary conditions in our numerical constructions of the black hole, thermal AdS, and wormholes solutions in sections \ref{sec:theoryandPD} and \ref{sec:ell0}. We work with a compact coordinate $y$ and Write $f(r)$ and $g(r)$ in a form adapted to the  asymptotically AdS boundary conditions \eqref{eq:bdymet}.

\subsection{Black hole solutions}

To construct black hole solutions, we choose our gauge so that
 \begin{equation}
 p(r)=r^2\,,\quad r\ge r_+\,,
 \end{equation}
 and require that the thermal circle shrinks to zero size at $r=r_+$. We thus write

 \begin{equation}
     f(r)=\left(\frac{r^2}{L^2}-\frac{r_+^2}{L^2}\right)q_1(r)\,,\quad 
 g(r)=\left(\frac{r^2}{L^2}-\frac{r_+^2}{L^2}\right)q_2(r)\,,\quad 
 \phi(r)=q_3(r)\left(1-\frac{r_+}{r}\right)^{-n/2}\,.
 \end{equation}

We will use a compact coordinate $y$ that maps the semi-infinite strip $r\geq r_+$ to the unit interval via
\begin{equation}\label{eq:bhy}
y=\frac{r-r_+}{r}\quad \Leftrightarrow \quad r = \frac{r_+}{1-y}\,.
\end{equation}
The conformal boundary is located at $y=1$, while the black hole horizon is mapped to $y=0$. We promote $r_+$ to be a function $r_+=Lq_4(y)$, and add a single equation
\begin{equation}
    \ddot q_4=0\,,
\end{equation}
where $\dot{}$ denotes differentiation with respect to $y$.  
The boundary conditions now read
\begin{subequations}
\begin{equation}\label{eq:bc1}
q_1(1)=q_2(1)=1\,,\quad q_3(1)=V\,,\quad \dot q_3(1)=V/2\,,\quad \text{and}\quad \dot q_4(1)=0
\end{equation}
and 
\begin{equation}
q_2(0)=\frac{1+3q_4(0)^2}{2q_4(0)^2}\,,q_1(0)=\frac{8\pi^2}{\beta^2[1+3q_4(0)^2]}\,,\quad q_3(0)=0\,,
\end{equation}
\end{subequations}%
The last boundary condition in \ref{eq:bc1} implies that $q_4$ is a constant.

To solve the resulting system of ordinary differential equations we discretize the resulting equations of motion for $q_1$, $q_2$, $q_3$, and $q_4$ using a single Chebyshev grid on Gauss-Lobatto nodes. The resulting system of nonlinear algebraic equations of motion is then solved using a standard Newton-Raphson routine. Since we seek solutions that are smooth in our integration domain  weexpect our discretization scheme to lead to exponential convergence. We use a similar procedure to construct the thermal AdS and wormhole solutions discussed below.

\subsection{Thermal AdS solutions}
 %%%%%%%%%%%%%%%%%%%%
 We now impose the gauge
 \begin{equation}
 p(r)=r^2
 \end{equation}
 and require that the $2$-sphere shrinks to zero size at $r=0$. We further define

 \begin{equation}
     f(r)=\left(\frac{r^2}{L^2}+1\right)q_1(r)\,,\quad
 g(r)=\left(\frac{r^2}{L^2}+1\right)q_2(r)\,,\quad
 \phi(r)=q_3(r)\,.
 \end{equation}
  The boundary conditions now read
 \begin{equation}
\lim_{r\to+\infty}q_1(r)=\lim_{r\to+\infty}q_2(r)=1\,,\quad \lim_{r\to+\infty}q_3(r)=V\,.
 \end{equation}
 At the origin, we further demand
 \begin{equation}
 q_2(0)=1\,,\quad  q_1^\prime(0)=q_3^\prime(0)=0\,,
 \end{equation}
where $^\prime$ denotes differentiation with respect to $r$.
 
We again use a compact coordinate $y$ that maps the semi-infinite strip $r\geq0$ to the unit interval via
\begin{equation}\label{eq:thy}
y=\frac{r}{r+L}\quad \Leftrightarrow\quad  r = \frac{Ly}{1-y}\,,
\end{equation}
with the conformal boundary being located at $y=1$, whereas the origin is mapped to $y=0$. The boundary conditions now read
\begin{subequations}
\begin{equation}
q_1(1)=q_2(1)=1\quad\text{and}\quad q_3(1)=V
\end{equation}
and
\begin{equation}
\dot{q}_1(0)=0\,,\quad q_2(0)=1\,,\quad q_3'(0)=0\,,
\end{equation}
\end{subequations}%
where $\dot{}$ denotes differentiation with respect to $y$. 

\subsection{Wormholes}\label{app:conwh}
 %%%%%%%%%%%%%%%%%%%%
To construct wormhole solutions, we impose the gauge 
 \begin{equation}
 p(r)=r^2+r_0^2.
 \end{equation}
We also require that the wormhole neck is located at $r=0$, about which we impose a $\mathbb{Z}_2$ symmetry $r\to-r$. We further define

 \begin{equation}
 f(r)=\left(\frac{r^2}{L^2}+\frac{r_0^2}{L^2}\right)q_1(r)\,,\quad
 g(r)=\left(\frac{r^2}{L^2}+\frac{r_0^2}{L^2}\right)q_2(r)\,,\quad
 \phi(r)=q_3(r)\,.
 \end{equation}
 The boundary conditions now read
 
 \begin{equation} \lim_{r\to+\infty}q_1(r)=\lim_{r\to+\infty}q_2(r)=1\,,
 \lim_{r\to+\infty}q_3(r)=V\,.
 \end{equation}
 At the wormhole neck, we further demand
 \begin{equation}
 q_1^\prime(0)=q_2^\prime(0)=q_3^\prime(0)=0\,,
 \end{equation}
 where $^\prime$ denotes differentiation with respect to $r$.
 
We use a compact coordinate $y$ that maps the semi-infinite strip $r\geq0$ to the unit interval via
\begin{equation}\label{eq:why}
y=\frac{r}{r+r_0}\quad\Leftrightarrow\quad r = \frac{y\,r_0}{1-y}\,,
\end{equation}
with the conformal boundary being located at $y=1$, whereas the wormhole neck is mapped to $y=0$. The boundary conditions now read
\begin{subequations}
\begin{equation}
q_1(1)=q_2(1)=1\quad\text{and}\quad q_3(1)=V
\end{equation}
and
\begin{equation}
\dot{q}_1(0)=\dot{q}_2(0)=\dot{q}_3(0)=0\,,
\end{equation}
\end{subequations}%
where $\dot{}$ represents differentiation with respect to $y$. 
To proceed, we promote $r_0\equiv Lq_4(y)$ to be a function of $y$, and add a single equation
\begin{equation}
\ddot{q}_4(y)=0\,.
\end{equation}
We then impose two more boundary conditions
\begin{equation}
\dot{q}_4(1)=0\,,\quad \mathrm{and} \quad 
q_1(0)=\left(\frac{2 \pi  n}{\beta }\right)^2 \frac{q_3(0)^2}{1-\ell  (1+\ell ) q_3(0)^2+3 q_4(0)^2}\,,
\end{equation}
with the latter coming from the reflection symmetry.

\section{Evaluation of the action and mass} \label{App:actionev}
This appendix describes technical details regarding the numerically stable evaluaton of the action for the solutions constructed above. We also provide the expressions for the masses of the solutions.

\subsection{Black holes}
Taking the trace of both sides of equation \eqref{eq:EinsteinEq} yields
\begin{equation}
    R=2\nabla_a\vec\Phi\cdot \nabla^a\vec\Phi^*\,.
\end{equation}
Inserting this into the action \eqref{eq:action1}, we obtain
\begin{equation}\label{eq:action3}
    16\pi GS=\int_\mathcal{M}\mathrm{d}^4 x\sqrt{g}\frac{6}{L^2}-2\int_{\partial\mathcal{M}}\mathrm{d}^3x\sqrt{h}K+S_{\partial\mathcal{M}}\,.
\end{equation}

To evaluate this action numerically, we again use the compact coordinate $y$ defined in \eqref{eq:bhy}. The metric now reads
\begin{equation}
    \mathrm{d}s^2=\frac{1}{(1-y)^2}\left[\frac{r_+^2(2-y)yq_1(y)}{L^2}\mathrm{d}\tau^2+\frac{L^2}{yq_2(y)(2-y)}\mathrm{d}y^2+r_+^2\mathrm{d}\Omega_2^2\right]\,,
\end{equation}
where $y\in[0,1]$.
The bulk term in the action \eqref{eq:action3} is manifestly divergent because $\sqrt{g}\sim (1-y)^{-4}$ as $y\to 1$.  However, this divergence is precisely canceled by the Gibbons-Hawking-York term and the counter terms. As a result, we can write the action in the form
\begin{equation}
\label{eq:evaluateactionbh}
    \frac{16\pi G S}{\beta L}=\frac{24\pi r_+^3}{L^2}\int_0^1 \mathrm{d}y \,\left[\frac{1}{(1-y)^4}\sqrt{\frac{q_1(y)}{q_2(y)}}-\mathcal{F}(y)\right]+\frac{24\pi r_+^3}{L^2}\int_0^1 \mathrm{d}y\mathcal{F}(y)-2\int_{\partial\mathcal{M}}\mathrm{d}^3x\sqrt{h}K+S_{\partial\mathcal{M}}\,,
\end{equation}
where we have included as a regulator the auxiliary function
\begin{equation}
\begin{split}
        \mathcal{F}(y)=&\frac{1}{(y-1)^4}+\frac{L^4 V^2 \omega^2}{2r_+^2(1-y)^2},
\end{split}
\end{equation}
which has the same singularity structure as the bulk integrand in equation \eqref{eq:action3} but which can be integrated analytically. Now the integrand in the first term on the right hand side of the equation above is finite everywhere, and the remaining three terms combine to give a finite expression. As a result, the black hole action is given by
\begin{equation}\label{eq:bhS}
\begin{split}
        \frac{16\pi G S}{\beta L}=& \frac{24\pi r_+^3}{L^2}\int_0^1 \mathrm{d}y \,\left[\frac{1}{(1-y)^4}\sqrt{\frac{q_1(y)}{q_2(y)}}-\mathcal{F}(y)\right]\\&+\frac{2 \pi  r_+\left(\left(q_1'''(1)-12\right) r_+^2-18 L^4 V^2 \omega ^2\right)}{3 L^2}\,.
\end{split}
\end{equation}

%%%%%%%%%%%%%%%%%%%%

\subsection{Thermal AdS}
We follow the same procedure as in the previous subsection. In terms of the compact coordinate $y$ defined in \eqref{eq:thy}, the metric reads
\begin{equation}
    \mathrm{d}s^2=\frac{1}{(1-y)^2}\left[(1+2y^2-2y)q_1(y)\mathrm{d}\tau^2+\frac{L^2}{1+2y^2-2y}\cdot\frac{1}{q_2(y)}\mathrm{d}y^2+L^2 y^2\mathrm{d}\Omega_2^2\right]\,.
\end{equation}
The action is evaluated via
\begin{equation}
\label{eq:evaluateactionth}
    \frac{16\pi G S}{\beta L}=24\pi L\int_0^1 \mathrm{d}y \,\left[\frac{y^2}{(1-y)^4}\sqrt{\frac{q_1(y)}{q_2(y)}}-\mathcal{F}(y)\right]+24\pi L\int_0^1 \mathrm{d}y\mathcal{F}(y)-2\int_{\partial\mathcal{M}}\mathrm{d}^3x\sqrt{h}K+S_{\partial\mathcal{M}}\,.
\end{equation}
Here we have included an auxiliary function
\begin{equation}
\begin{split}
        \mathcal{F}(y)=&\frac{1}{(y-1)^4}-\frac{2}{(1-y)^3}+\frac{L^2 V^2 \omega ^2+2}{2 (1-y)^2}
\end{split}
\end{equation}
so that the second integral in equation \eqref{eq:evaluateactionth} can be evaluated analytically while making the first integral manifestly finite. As a result, we find
\begin{equation}\label{eq:thS}
\begin{split}
        \frac{16\pi G S}{\beta L}=& 24\pi L\int_0^1 \mathrm{d}y \,\left[\frac{y^2}{(1-y)^4}\sqrt{\frac{q_1(y)}{q_2(y)}}-\mathcal{F}(y)\right]\\&+\frac{2}{3} \pi  L \left(q_1'''(1)-12 V^2 \left(-2 L^2 \omega ^2+\ell ^2+\ell \right)\right)\,.
\end{split}
\end{equation}

\subsection{Wormholes}\label{app:wham}

We again proceed much as above.
Using the compact coordinate $y$ defined in \eqref{eq:why}, the metric is
\begin{equation}
    \mathrm{d}s^2=\frac{1}{(1-y)^2}\left[\frac{r_0^2}{L^2}(1+2y^2-2y)q_1(y)\mathrm{d}\tau^2+\frac{L^2}{1+2y^2-2y}\cdot\frac{1}{q_2(y)}\mathrm{d}y^2+r_0^2(1+2y^2-2y)\mathrm{d}\Omega_2^2\right]\,.
\end{equation}
Here $y\in[0,1]$, so that our solution represents half of the ${\mathbb Z}_2$-symmetric wormhole geometry or, equivalently, the ${\mathbb Z}_2$-quotient of our wormhoe. The action of this quotient is thus given by
\begin{equation}
\label{eq:evaluateactionwh}
    \frac{16\pi G S}{\beta L}=\frac{24\pi r_0^3}{L^2}\int_0^1 \mathrm{d}y \,\left[\frac{1+2y^2-2y}{(1-y)^4}\sqrt{\frac{q_1(y)}{q_2(y)}}-\mathcal{F}(y)\right]+\frac{24\pi r_0^3}{L^2}\int_0^1 \mathrm{d}y\mathcal{F}(y)-2\int_{\partial\mathcal{M}}\mathrm{d}^3x\sqrt{h}K+S_{\partial\mathcal{M}}\,.
\end{equation}
Here the auxiliary function is
\begin{equation}
\begin{split}
        \mathcal{F}(y)=&\frac{1}{(y-1)^4}-\frac{2}{(1-y)^3}+\frac{\frac{L^4 V^2 \omega ^2}{r_0^2}+3}{2 (1-y)^2}\,.
\end{split}
\end{equation}
The final expression for the action of a half-wormhole is thus
\begin{equation}\label{eq:whS}
\begin{split}
        \frac{16\pi G S}{\beta L}=& \frac{24\pi r_0^3}{L^2}\int_0^1 \mathrm{d}y \,\left[\frac{1+2y^2-2y}{(1-y)^4}\sqrt{\frac{q_1(y)}{q_2(y)}}-\mathcal{F}(y)\right]\\&+\frac{2 \pi  \left(q_2'''(1)+12\right) r_0^3}{3 L^2}+8 \pi  L^2 r_0 V^2 \omega ^2-8 \pi  r_0 \left(V^2 \ell  (\ell +1)-1\right)\,.
\end{split}
\end{equation}
The action of the full wormhole geometry is then obtained by multiplying by 2. However, in the main text we simply use \eqref{eq:whS} and compare it with the action for a single (1-boundary) black hole or thermal AdS solution.  

\subsection{Wormhole Mass}
\label{app:wmass}

We also need to compute the `mass' of our wormholes (though we do not make use of this quantity for other phases).  By `mass,' we mean here the integral of the $\tau\tau$ component of the boundary stress tensor (multiplied by the square root of the appropriate induced metric) over a boundary hypersurface of constant $\tau$.  

This quantity is easiest to compute by writing the compact radial coordinate $y$ in terms of the appropriate Fefferman-Graham coordinate $z$.   In particular, for wormholes in a theory of Einstein gravity with two sets of scalars, one with angular momentum number $\ell$ and boundary value $V_\ell$, the other with angular momentum number $0$ and boundary value $V_0$, near the boundary we find
\begin{equation}
    y=1+r_0z+r_0^2 z^2+\frac{r_0}{4}\left[1+6r_0^2-V_\ell^2\ell(\ell+1)+(V_0^2+V_\ell^2)\omega^2\right]z^3+\mathcal{O}(z^4)\,.
\end{equation}  
 Taking $V_0=0$ gives us wormhole masses with a single set of scalar fields with angular momentum number $\ell$ that agree with those in section \ref{sec:EGCS}.
Since we consider the case of $d=3$ boundary dimensions, the `mass' of our wormhole is proportional to $g_{\tau\tau}^{[3]}$, which is the coefficient of the $z^1$ term in $g_{\tau\tau}$.  Using our ansatz and boundary conditions yields 
\begin{equation}
\label{eq:massEq}
\begin{split}
    \mathrm{mass}= \frac{2\pi}{3r_0^3}\Big\{&-12 r_0^6-12 r_0^4\left[-1+V_\ell^2\ell(\ell+1)-2(V_0^2+V_\ell^2)\omega^2\right]\\
    &+6r_0^4\left[1+6 r_0^2-V_\ell^2\ell(\ell+1)+(V_0^2+V_\ell^2)\omega^2\right]\\
    &+r_0^6\left[12-\frac{12(V_0^2+V_\ell^2)\omega^2}{r_0^2}+q_2'''(1)\right]\\
    &-\frac{r_0^4}{3}\left[30+84 r_0^2-30V_\ell^2\ell(\ell+1)+30(V_0^2+V_\ell^2)\omega^2+r_0^2q_2'''(1)\right]\Big\}\,,
\end{split}
\end{equation}
 We use this expression below with $\ell=1$ to study the behavior of our wormholes at fixed mass in section \ref{sec:ell0}.

\section{Euclidean Stability following the rule-of-thumb prescription}
\label{app:ROT}

This appendix studies the stability of our wormhole saddles using the rule-of-thumb framework described in \cite{Marolf:2022ntb}.  The rule-of-thumb approach generalizes the Wick-rotate-the-pure-trace-mode prescription of \cite{Gibbons:1978ac}. We will simply apply the method here, referring the reader to \cite{Marolf:2022ntb} for motivations and a full description of the procedure, and to \cite{Liu:2023jvm} for discussion of subtleties.

As in section \ref{sec:stability}, it is useful to write the complex scalar field in terms of a pair of real scalars which we call the real and imaginary parts.  This avoids various confusions associated with factors of $i$\, that appear due to our use of complex exponential Fourier modes . We denote the linear space of the field perturbations $(h_{ab},\vec\psi_R,\vec\psi_I)$ by $\mathfrak{G}$:
\begin{equation}
    \mathfrak{G}=\{\mathfrak{h}|\mathfrak{h}=(h_{ab},\vec{\psi}_R,\vec{\psi}_I)\}\,,
\end{equation}
where $\vec{\psi}_R$ and $\vec{\psi}_I$ are $(2\ell+1)$ dimensional real vectors so that the complex-valued perturbation is $\vec\psi=\vec{\psi}_R+i\vec{\psi}_I$\,. We impose the following inner product on $\mathfrak{G}$:
\begin{equation}
    (h,\vec\psi_R,\vec{\psi}_I;\tilde h,\vec{\tilde \psi}_R,\vec{\tilde\psi}_I)_{{\mathcal{G}}}=\frac{1}{32\pi G}\int_{\mathcal{M}}\dd ^4x\,\sqrt{\hat g} \,(h_{ab}{\mathcal{G}}^{abcd}_{-1}\tilde h_{cd}+\vec \psi_R\cdot \vec{\tilde \psi}_R+\vec \psi_I\cdot \vec{\tilde \psi}_I)\,,
\end{equation}
where
\begin{equation}
{\mathcal{G}}^{abcd}_{-1}=\frac{1}{2}(g^{ac} g^{bd}+ g^{ad} g^{bc}- g^{ab} g^{bd})
\end{equation}
is the DeWitt$_{-1}$ metric built from the background metric $g_{ab}$. 

The quadratic action can be written
\begin{equation}
    S^{[2]}=(\mathfrak{h},\mathcal{L}\mathfrak{h})_{\mathcal{G}}\,.
\end{equation}
Due to gauge symmetry of our gravitational system, any fluctuation operator $\mathcal{L}:\mathfrak{G}\to\mathfrak{G}$ will be highly degenerate at eigenvalue $\lambda=0$. In particular, the above quadratic action is invariant under
\begin{equation}
    h_{ab}\to h_{ab}+\nabla_{(a}\xi_{b)}\,,\quad \vec{\psi}\to \vec{\psi}+\xi^a\nabla_a\vec\Phi_R+i\xi^a\nabla_a\vec{\Phi}_I\,,
\end{equation}
where $\vec\Phi_{R/I}$ is the real/complex part of the background scalar field $\Phi$\,. As a result, for appropriate real vector fields $\xi^a$, $\mathcal{L}$ must annihilate any pure-gauge mode $\mathfrak{h}=(\nabla_{(a}\xi_{b)},\xi^a\nabla_a\vec\Phi_R,\xi^a\nabla_a\vec{\Phi}_I)$\,.

Given a metric on the space of perturbations, it is thus natural to attempt to choose a gauge condition that is
satisfied precisely by the space $W^{\perp}$ of perturbations that are orthogonal to the space $W$ spanned by pure-gauge
modes
The construction of the pure-gauge mode~$\mathfrak{h}=(\nabla_{(a}\xi_{b)},\xi^a\nabla_a\vec\Phi_R,\xi^a\nabla_a\vec{\Phi}_I)$ from the vector field $\xi^a$ can be described by a linear map $\mathcal{P}:V\to W\subset \mathfrak{G}$, where $V$ is the space of smooth vector fields with $\xi(1)=\xi'(1)=0$, and where $P\xi=(\nabla_{(a}\xi_{b)},\xi^a\nabla_a\vec\Phi_R,\xi^a\nabla_a\vec{\Phi}_I)$\,. Introducing a positive-definite Hermitian inner product $(\xi,\tilde\xi)_V$ on $V$,
\begin{equation}
    (\xi,\tilde\xi)_V=\frac{1}{32\pi G}\int_{\mathcal{M}}\dd^4 x\,\sqrt{g}\, g_{ab}\xi^{a} \tilde \xi^b\,,
\end{equation}
we can define the adjoint operator of $\mathcal{P}$ by requiring
\begin{equation}
    (\mathfrak{h},\mathcal{P}\xi)_{\mathcal{G}}=(\mathcal{P}^\dagger\mathfrak{h},\xi)_V\,,\quad \mathcal{P}^\dagger: \mathfrak{G}\to V\,.
\end{equation}
A short calculation then shows:
\begin{equation}
    \label{eq:gaugefixing}(\mathcal{P}^\dagger\mathfrak{h})_b=\nabla_b h-2\nabla^a h_{ab}+\vec\psi_R\nabla_a\vec{\Phi}_R+\vec\psi_I\nabla_a\vec{\Phi}_I\,.
\end{equation}
Following \cite{Liu:2023jvm}, we can take equation \eqref{eq:gaugefixing} as our gauge condition as long as the operator $G:=\mathcal{P}^\dagger\mathcal{P}$ is an invertible map from $V$ to itself. Note that
\begin{equation}
    (G\xi)_b=-2\left(\nabla^2\xi_b+\Lambda\xi_b\right)+\xi^a\nabla_a\vec{\Phi}_R\cdot\nabla_b\vec{\Phi}_R+\xi^a\nabla_a\vec{\Phi}_I\cdot\nabla_b\vec{\Phi}_I\,.
\end{equation}
Contracting this with $\sqrt{g}\xi^b$ and integrating over the spacetime $\mathcal{M}$ yields
\begin{equation}
    \int_{\mathcal{M}}\dd^4 x\,\sqrt{g}\left[2(\nabla^a\xi^b)(\nabla_a\xi_b)-2\Lambda \xi^b\xi_b+\left(\xi^a\nabla_a\vec{\Phi}_R\right)^2+\left(\xi^a\nabla_a\vec{\Phi}_I\right)^2 \right]\,,
\end{equation}
which is manifestly positive. We will thus choose our gauge condition to be equation \eqref{eq:gaugefixing}\,.

Note that $\vec\psi$ appears algebraically in equation~\eqref{eq:gaugefixing}.  As a result, we can impose the gauge condition explicitly by solving $\delta\phi$.  We then have no need to add further gauge-symmetry-breaking terms as in \cite{Marolf:2022ntb}.

Since this second method is intended as a check on the main results of section \ref{sec:stability}, and since we expect modes with non-zero $k,\ell_S$ to be more stable, we apply the rule-of-thumb method only to modes with $k=\ell_S=0.$   We again use the ansatz for perturbations given in equations \eqref{eq:pertm0g} and \eqref{eq:pertm0phi}\ and the associated quadratic action $S_A^{[2]}$ introduced in section \ref{sec:stability},. Imposing the gauge condition \eqref{eq:gaugefixing} by solving for $\delta q_3$ yields 
\begin{equation}
\begin{split}
     \delta q_3(y)=&\frac{1}{(-1+y) \left(2 y^2-2 y+1\right) q_1(y)q_3'(y)}\cdot\\\Big\{
     &2y\delta q_2(y)+4yq_1(y)\delta q_4(y)+(-1+y)(1-2y+2y^2)\left[q_2(y)\delta q_2(y)q_1'(y)-\delta q_1'(y)\right]\\+&q_1(y)q_2(y)\left[-6y\delta q_2(y)+(-1+y)(1-2y+2y^2)\delta q2'(y)\right]\\
     +&q_1(y)(-1+y)(1-2y+2y^2)(\delta q_2(y)q2'(y)-2\delta q_4'(y))
    \Big\}\,.
\end{split}
\end{equation}
Inserting this expression into $S_A^{[2]}$ then gives a gauge-fixed quadratic action built from $\delta q_1\,,\delta q_2\,,\delta q_4$ and their derivatives. The boundary conditions for $\delta q_3$ now impose new boundary conditions for $\delta q_1\,,\delta q_2\,,\delta q_4$ which take the form
\begin{equation}
    \delta q_2''(1)=0\,,\quad 6\delta q_1''(1)+12\delta q_4''(1)+\delta q_1^{(3)}(1)+3\delta q_2^{(3)}(1)+2 \delta q_4^{(3)}(1)=0\,,
\end{equation}
and 
\begin{equation}
    -6\delta q_1''(1)+\delta q_1^{(3)}(0)+L^2\kappa^2\omega\left[\frac{6\delta q_2''(0)-\delta q_2^{(3)}(0)}{y_0^2}+\frac{2(-6\delta q_4''(0)+\delta q_4^{(3)}(0))}{1+3y_0^2-\ell(\ell+1)\kappa^2}\right]=0
\end{equation}
in the even sector, or 
\begin{equation}
    \begin{split}
        \delta q_1'(0)+L^2\kappa^2\omega^2\left[-\frac{\delta q_2'(0)}{y_0^2}+\frac{2\delta q_4'(0)}{1+3y_0^2-\ell(\ell+1)\kappa^2}\right]=0\,,\\
        -3\delta q_1'(0)+\delta q_1''(0)+L^2\kappa^2\omega\left[\frac{3\delta q_2'(0)-\delta q_2''(0)}{y_0^2}+\frac{-6\delta q_4'(0)+2\delta q_4''(0)}{1+3y_0^2-\ell(\ell+1)\kappa^2}\right]=0
    \end{split}
\end{equation}
in the odd sector, where $\kappa=q_3(0)$ is the scalar field at the wormhole neck.

\begin{figure}
    \centering
    \includegraphics[width=0.98\linewidth]{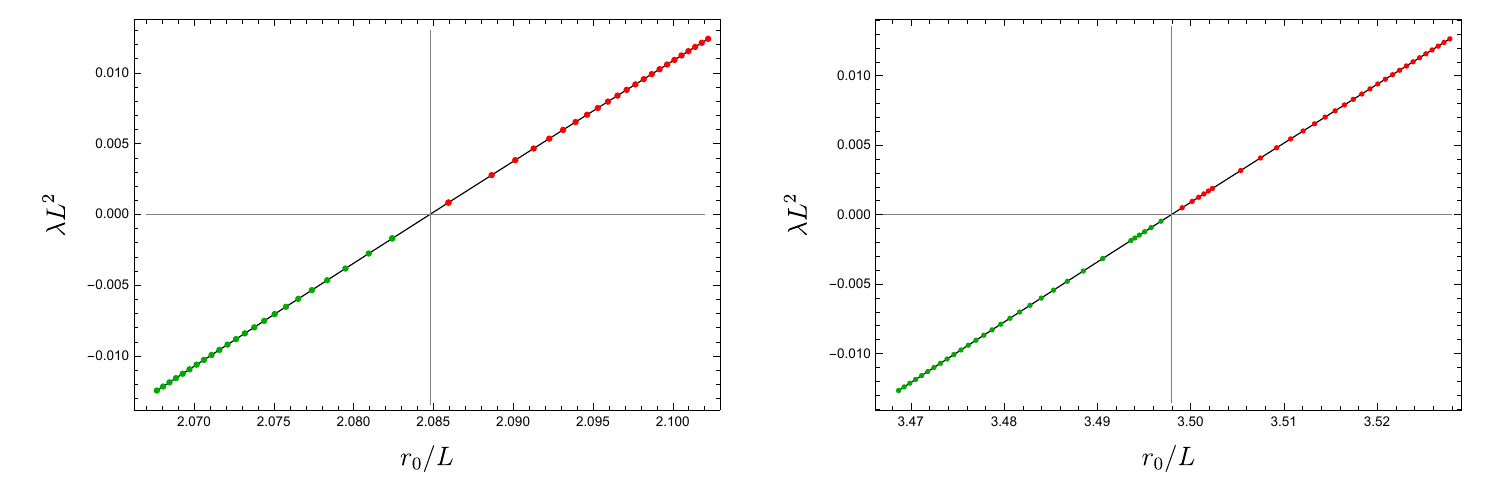}
    \caption{The lowest lying mode of $\mathcal{L}$ using  the rule-of-thumb prescription with the DeWitt$_{-1}$ metric.  The left/right panels show $\ell=1$ and $\ell=2$ with $n=\beta=1$. The horizontal lines are $\lambda=0$ while the vertical lines mark the critical value of $r_0$ from figure~\ref{fig:fxibeta}. The $\lambda<0$ green dots to the left are thus small wormholes while the $\lambda >0$ red dots to the right are large wormholes.}
    \label{fig:spec_actionmethod}
\end{figure}

We now analyze the spectrum of the operator ${\mathcal L}$ defined by the above action and the inner product $(,)_{\mathcal{G}}$ using numerical methods analogous to those in section \ref{sec:stability} 
We find a single negative mode in the even sector, and no negative modes in the odd sector.  The lowest lying mode is given in figure \ref{fig:spec_actionmethod}.

These results thus agree with those of section \ref{sec:stability}. 

\section{Perturbative stability of wormholes in the non-static or non-spherical sectors}\label{app:sectors}
\subsection{The non-static spherical sector} %The perturbed scalar can be written as
%\begin{equation}
%    \phi_{m}(\tau,y,\theta,\varphi)=\Phi(y,\tau)e^{-i\frac{2\pi n}{\beta L}\tau}\times \cdots\,,\quad\mathrm{and}\quad \phi^*_{m}(\tau,y,\theta,\varphi)=\Phi^*(y,\tau)e^{-i\frac{2\pi n}{\beta L}\tau}\times \cdots\,.
%\end{equation}
In the non-static sector, it turns out to be useful to use an ansatz that differs slightly from the one in the static spherical sector used in section \ref{sec:poaformalism}.
In parallel with our method of finding the wormhole solutions explained in previous appendices, we now take 
perturbed fields to have the following form:  
\begin{equation}\label{eq:perttau}
    \begin{split}
        g_{\tau\tau}=\frac{r_0^2(1+2y^2-2y)q_1(y)}{(1-y)^2 L^2}+\epsilon\, \delta q_1(y,\tau)\,,&\quad g_{\theta\theta}=\frac{r_0^2(1+2y^2-2y)}{(1-y)^2}+\epsilon\, \delta q_4(y,\tau)\,,\\
        \beta=0+\epsilon\,\frac{\delta\chi(y,\tau)}{(1-y)^2}\,,&\quad \alpha=\sqrt{\frac{L^2}{q_2(y)(1-y)^2(1+2y^2-2y)}}+\epsilon\,\delta q_2(y,\tau)\,,\\
        \phi=q_3(y)+\epsilon\,\delta q_3(y,\tau)\,,&\quad \phi^*=q_3(y)+\epsilon\,\delta q_3^*(y,\tau)\,.\\
    \end{split}
\end{equation}
Note that $\delta q_1, \delta q_2$ are not just variations of $q_1,q_2$).

%One can prove that, although the scalar field breaks the U$(1)$ symmetry on the thermal-$\tau$ circle, different Fourier modes with different $m$ are still decoupled from each other. As a result, we can work in a sector that has a single Fourier mode number $m$\,.

It is useful to write each complex scalar in terms of a pair of real scalars and to then consider only real perturbations of these latter scalars.  Since any such real perturbation will include mode numbers $k$ and $-k$ on the $S^1$ with complex-conjugate Fourier coefficients, the (real) metric perturbations take the form
\begin{equation}
    \delta q_1(y,\tau)=(\delta q_{1,R}+i\delta q_{1,I})e^{i \frac{2\pi k}{\beta\,L} \tau}+(\delta q_{1,R}-i\delta q_{1,I})e^{-i \frac{2\pi k}{\beta\,L} \tau}\,
\end{equation}
for real $\delta q_{1,R},\delta q_{1,I}$ (and
similarly for $\delta\chi,\delta q_2,\delta q_4$\,).  For the (complex) scalars, we write
\begin{equation}
    \begin{split}
        \delta q_3(y,\tau)&=(\delta q_{3,R}+i\delta q_{3,I})e^{i  \frac{2\pi k}{\beta\,L} \tau}+(\delta Q_{3,R}-i\delta Q_{3,I})e^{-i  \frac{2\pi k}{\beta\,L} \tau}\,,\\
        \delta q_3^*(y,\tau)&=(\delta q_{3,R}-i\delta q_{3,I})e^{-i  \frac{2\pi k}{\beta\,L} \tau}+(\delta Q_{3,R}+i\delta Q_{3,I})e^{i  \frac{2\pi k}{\beta\,L} \tau}\,.
    \end{split}
\end{equation}
For later use, we note that
the most general infinitesimal diffeomorphism with angular momentum $k$ on the $S^1$ takes the form 
\begin{equation}
\xi=L^2[(\xi_{0,r}+i\xi_{0,i})e^{i \frac{2\pi k}{\beta\,L} \tau}+(\xi_{0,r}-i\xi_{0,i})e^{-i \frac{2\pi k}{\beta\,L} \tau}]\dd \tau+L^2[(\xi_{1,r}+i\xi_{1,i})e^{i \frac{2\pi k}{\beta\,L} \tau}+(\xi_{1,r}-i\xi_{1,i})e^{-i \frac{2\pi k}{\beta\,L} \tau}]\dd y\,.
\end{equation}

We now manipulate the action and introduce gauge-invariant variables in direct analogy with our procedure as in the static spherical sector, though with more variables. 
After writing the quadratic action in terms of first derivatives, we can perform further integrations by parts until  $\delta \chi_I,\delta\chi_R,\delta q_{2,I},\delta q_{2,R}$ all appear in the quadratic action algebraically.  This is again a result of Bianchi identities, and it related to the fact that these are modes of the linearized radial lapse $(\delta q_{2,R},\delta q_{2,I})$ and shift $(\delta \chi_{2,R},\delta \chi_{2,I})$. We then again choose the contour so that we may perform the Gaussian integral over these four variables away from the ${\mathbb Z}_2$-invariant neck at $y=0$, and so that integration over $(\delta q_{2,R},\delta q_{2,I})$ at $y=0$ imposes the appropriate linearized constraint.  (There is no corresponding issue for the linearized shift variables $\delta \chi_{R,I}$.)  The result is a new quadratic action expressed in terms of $\{\delta q_{1,I/R}\,,\delta q_{3,I/R},\delta Q_{3,I/R}\,,\delta q_{4,I/R}\}$ and their derivatives, and which is again free of divergences. We then introduce four gauge invariant variables:
\begin{equation}
    \begin{split}
        P_I=\delta q_{1,I}+A_1 \delta Q_{3,I}+B_1\delta q_{4,I}\,,\quad 
        Q_I=\delta Q_{3,I}+A_2 \delta q_{3,I}+B_2\delta q_{4,I}\,,\\
        P_R=\delta q_{1,R}+A_1 \delta Q_{3,R}+B_1\delta q_{4,R}\,,\quad 
        Q_R=\delta Q_{3,R}+A_2 \delta q_{3,R}+B_2\delta q_{4,R}\,,\\
    \end{split}
\end{equation}
where $A_1,A_2,B_1,B_2$ are determined by the background fields according to
\begin{equation}
\begin{split}
     A_1&=-\frac{2 \tilde{k} r_0^2(1-2y+2y^2)q_1(y)}{L^2(1-y)^2\omega q_3(y)}\,,\\
     A_2&=1\,,\\
     B_1&=\frac{ (y-1) (1-2y+2y^2) q_1'(y)-2 y q_1(y)}{2 L^2 y }+\frac{ \tilde{k} (1-y) (1-2y+2y^2) q_1(y) q_3'(y)}{ L^2 y \omega  q_3(y)}\,,\\
     B_2&=-\frac{(1-y)^3q_3'(y)}{r_0^2 y}\,,
\end{split}
\end{equation}
where $\tilde k=2\pi k/(\beta L)$.  Notice the similarity of $P_{I/R},Q_{I/R}$ in the above definitions. Further algebra and integrations by parts then allow use to express our quadratic action in terms of only $\{P_{I/R},Q_{I/R}\}$, their derivatives, and the background fields. Furthermore, we find that $\{P_I,Q_I\}$ and $\{P_R,Q_R\}$ decouple, i.e.,
\begin{equation}    \hat{S}^{[2]}=\hat{S}^{[2]}_I(P_I,P_I',Q_I,Q_I')+\hat{S}^{[2]}_R(P_R,P_R',Q_R,Q_R')\,,
\end{equation}
where $^\prime$ denotes derivative with respect to $y$, and where the functional forms of $\hat{S}^{[2]}_I$ and $\hat{S}^{[2]}_R$ are identical (as they are related by the action of our $U(1)$ symmetry that turns the appropriate $\sin$ functions into $\cos$ functions).  As a result, it suffices to analyze only the spectrum of, say, the $I$ modes since the spectrum of the $R$ modes will be identical, i.e., the two sectors are isospectral.   Using an inner product analogous to \eqref{eq:IP1} to turn second derivatives of 
$\hat{S}^{[2]}$ intro a self-adjoint linear operator $\mathcal{L}$, we find no negative modes with $k\neq 0$ for any wormhole.

\subsection{Static non-spherical scalar-derived deformations with $\ell_S\geq2$}
Let us define the $(2\,\ell+1)$-component vector
\begin{equation}
\begin{aligned}
&\left(\vec{X}\right)_m=\sqrt{2}\sqrt{\frac{(\ell-m)!}{(\ell+m)!}}P^{\ell}_m(\cos\theta)\sin(m\varphi)\,\quad -\ell\leq m\leq-1
\\
&\left(\vec{X}\right)_0=P^{\ell}(\cos\theta)\,\quad
\\
&\left(\vec{X}\right)_m=\sqrt{2}\sqrt{\frac{(\ell-m)!}{(\ell+m)!}}P^{\ell}_m(\cos\theta)\cos(m\varphi)\,\quad 1\leq m\leq \ell\,,
\end{aligned}
\end{equation}
so that the background scalar field in Eq.~(\ref{eq:phiansatz}) can be written as
\begin{equation}
\vec{\phi}=\phi(r)e^{-i\frac{2\pi\,n}{\beta\,L}\tau}\vec{X}\,.
\end{equation}

We adopt the following Ansatz for the field-theoretic negative modes
\begin{subequations}
\begin{equation}
\delta {\rm d}s^2= \left[h_{\tau\tau}(r){\rm d}\tau^2+h_{rr}(r){\rm d}r^2+h_L(r){\rm d}\Omega_2^2\right]\mathbb{Y}^{m_S\,\ell_S}+2 h_r(r){\rm d}r\,{\rm d}y^i\Grad_i\mathbb{Y}^{m_S\,\ell_S}+h_T(r){\rm d}y^i\mathbb{Y}^{m_S\,\ell_S}_{ij}{\rm d}y^j
\label{eq:scalarstaticm}
\end{equation}
with $i = 1,2$ parameterizing coordinates on the unit-radius $S^2$, and where $\Grad$ denotes the affine connection on $S^2$. The traceless two-tensor $\mathbb{Y}^{m_S\,\ell_S}_{ij}$ is defined as
\begin{equation}
\mathbb{Y}^{m_S\,\ell_S}_{ij}=\Grad_i\Grad_j\mathbb{Y}^{m_S\,\ell_S}+\frac{\ell_S(\ell_S+1)}{2}\mathbb{G}_{ij}\mathbb{Y}^{m_S\,\ell_S}\,,
\end{equation}
\end{subequations}%
where $\mathbb{G}_{ij}$ are the components of the metric on a unit-radius round two-sphere. The $\mathbb{Y}^{m_S\,\ell_S}$ are the standard scalar spherical harmonics of degree $\ell_S$ and order $m_S$, satisfying
\begin{equation}
\DD \mathbb{Y}^{m_S\,\ell_S}+\ell_S(\ell_S+1)\mathbb{Y}^{m_S\,\ell_S}=0\,,
\end{equation}
where $\DD$ denotes the Laplacian operator on $S^2$. Note also that $|m_S|\leq \ell_S$.

For the scalar field, we take
\begin{equation}
\delta\vec{\phi}(\tau,r,\theta,\varphi)=e^{-i\frac{2\pi\,n}{\beta\,L}\tau}\left[\phi_0(r)\vec{X}+\sum_{I=1}^{\ell}\phi_I(r)(\underbrace{\Grad\ldots \Grad}_{I})_{i_1\ldots i_I}\mathbb{Y}^{m_S\,\ell_S}(\underbrace{\Grad\ldots \Grad}_{I})^{i_1\ldots i_I}\,\vec{X}\right]\,,
\label{eq:scalarstatics}
\end{equation}
so that a generic deformation is parametrised by $(\ell+1)$ perturbations $\{\delta \phi_0,\delta \phi_I\}$, with $I=1,\ldots,\ell$. We have explicitly checked that the above Ansatz for the scalar deformations allows us to study generic \emph{on-shell} perturbations for $\ell = 0, \ldots, 4$, and we believe that it will work for any value of $\ell$. Note that for a fixed value of $\ell$ (which fixes a given saddle) we can take $\ell_S\geq2$. The case with $\ell_S=0$ was covered in the previous section, and the case with $\ell_S=1$ has to be taken care off separately, since $\mathbb{Y}^{m_S\,1}$ vanishes identically for $|m_S|\leq1$.

Generic infinitesimal diffeomorphisms consistent with staticity, can be written as
\begin{equation}
\xi_a{\rm d}x^a=\xi_{r}(r)\mathbb{Y}^{m_S\,\ell_S}\,{\rm d}r+z(r){\rm d}y^i\Grad_i\mathbb{Y}^{m_S\,\ell_S}\,.
\label{eq:gauparam}
\end{equation}
Under such transformations, the metric and scalar field components transform as
\begin{equation}
\begin{aligned}
&\delta h_{\tau\tau}(r)= g(r)\xi_r(r)f^\prime(r)\,,\quad \delta h_{rr}(r)=\frac{g^\prime(r)}{g(r)}\xi_r(r)+2 \xi_r^\prime(r)\,,
\\
&\delta f_r(r)=\xi_r(r)+z^\prime(r)-\frac{2 r\,z(r)}{r^2+r_0^2}\,,\quad \delta h_L(r)=2 r g(r)\xi_r(r)-\ell_S(\ell_S+1)z(r)\,,
\\
&\delta h_T(r)=2 z(r)\,,\quad \delta \phi_0(r)=g(r)\phi^\prime(r)\xi_r(r)\,,
\\
&
\delta\phi_1(r)=\frac{z(r)\phi(r)}{r^2+r_0^2}\,,\quad \text{and}\quad  \delta\phi_I(r)=0\,,\quad\text{for}\quad I\geq2\,.
\label{eq:gaugescalar}
\end{aligned}
\end{equation}

Since the action can be written entirely in terms of the Hamiltonian and momentum constraints\footnote{This requires a number of integration by parts, whose boundary terms can be shown to vanish with our choice of boundary conditions}, and because our metric deformations depend nontrivially only on $r$, the components $\{h_{ar},f_r\}$ appear algebraically in the quadratic action. For this reason, these can be readily integrated out. As in the previous section, we imagine deforming the integration contour appropriately to ensure convergence of the quadratic integral. Effectively, this procedure amounts to imposing the constraint equations on off-shell configurations. At this stage, we introduce $\ell + 2$ gauge-invariant quantities, denoted by $Q_I$ with $I = 1, \ldots,\ell+1$:
\begin{equation}
\begin{aligned}
&Q_1(r)=h_{\tau \tau }(r)-\frac{h_L(r) f'(r)}{2 r}-\frac{\ell _S \left(1+\ell _S\right) h_T(r) f'(r)}{4 r}\,,
\\
&Q_2(r)=\phi _0(r)-\frac{h_L(r) \Phi '(r)}{2 r}-\frac{\ell _S \left(1+\ell _S\right) h_T(r) \Phi '(r)}{4 r}\,,
\\
&Q_3(r)=\phi _1(r)-\frac{\Phi (r) h_T(r)}{2 \left(r^2+r_0^2\right)}\,,
\\
&Q_I(r)=\phi_I(r),\quad \text{for}\quad I\geq2\,.
\end{aligned}
\label{eqs:gaugescalark0}
\end{equation}
It is a straightforward, albeit tedious, exercise to solve for $\{h_{\tau\tau}, \phi_0, \ldots, \phi_{\ell}\}$ in terms of $\{Q_I, h_T, h_L\}$. After performing several integrations by parts, the dependence on $\{h_L,h_T\}$ drops out of the quadratic action, leaving us with a quadratic action that depends only on the $\ell + 2$ gauge-invariant quantities $Q_I$. This is the quadratic action whose positivity properties we wish the study. At this stage we proceed numerically.

\subsection{Static non-spherical scalar-derived deformations with $\ell_S=1$}
This sector is, in many respects, similar to the previous one, except that for perturbations with $\ell_S = 1$, the tensor harmonic $\mathbb{Y}^{m_S,1}_{ij}$ vanishes identically. As a result, $h_T$ does not appear in the metric \emph{Ansatz}. However, the gauge parameter $z(r)$ introduced in Eq.(\ref{eq:gauparam}) remains nonzero. The gauge transformations are exactly as in Eq.(\ref{eq:gaugescalar}) with $\ell_S = 1$, except that there is no transformation rule for $h_T$.

As before, we can integrate out $h_{ar}$, leaving a quadratic action that depends on $h_{\tau\tau}$, $\phi_0$, $\phi_1$, $\phi_I$, and $h_L$. At this point, we introduce gauge-invariant quantities:
\begin{equation}
\begin{aligned}
&Q_1(r)=\phi _0(r)-\frac{h_{\tau \tau }(r) \Phi '(r)}{f'(r)}\,,
\\
&Q_2(r)=\phi _1(r)+\frac{\Phi (r) h_L(r)}{2 \left(r^2+r_0^2\right)}-\frac{r \Phi (r) h_{\tau \tau }(r)}{\left(r^2+r_0^2\right) f'(r)}\,,
\\
&Q_I(r)=\phi_I(r),\quad \text{for}\quad I\geq2\,,
\end{aligned}
\end{equation}
which now differ slightly from those in Eqs.~(\ref{eqs:gaugescalark0}), since $h_T$ does not appear in the quadratic action. Its role is instead taken by $h_{\tau\tau}$. One can explicitly express $\{\phi_0, \ldots, \phi_{\ell}\}$ in terms of ${Q_I, h_{\tau\tau}, h_L}$, though the procedure is somewhat laborious. After performing several integrations by parts, all dependence on $\{h_L, h_{\tau\tau}\}$ cancels out of the quadratic action, which then involves only the $\ell + 1$ gauge-invariant quantities $Q_I$. This is the action whose positivity properties we aim to analyse. At this stage, we proceed numerically.

\subsection{Static non-spherical vector-derived deformations with $\ell_V\geq2$}
We adopt the following Ansatz for the metric and gauge field
\begin{equation}
\delta {\rm d}s^2=2f_r(r)\,\mathbb{Y}^{\ell_V\,m_V}_{i}\,{\rm d}r\,{\rm d}y^i+h_T(r)\left(\Grad_i\mathbb{Y}^{\ell_V\,m_V}_j+\Grad_j\mathbb{Y}^{\ell_V\,m_V}_i\right)\,{\rm d}y^i{\rm d}y^j
\label{eq:vec}
\end{equation}
with $\mathbb{Y}_i$ a vector spherical harmonic satisfying
\begin{equation}
\DD\mathbb{Y}^{\ell_V\,m_V}_i+\left[\ell_V(\ell_V+1)-1\right]\mathbb{Y}^{\ell_V\,m_V}_i=0\,,
\end{equation}
with $\ell_V\geq1$. In the special case of the two-sphere, there is a simple expression for $\mathbb{Y}^{\ell_V,m_V}_i$ in terms of the standard scalar spherical harmonics $\mathbb{Y}^{\ell_V,m_V}$ on $S^2$, namely:
\begin{equation}
\mathbb{Y}_i{\rm d}y^i={}_{2}\star\left({\rm d}\mathbb{Y}^{\ell_V\,m_V}\right)\,,
\end{equation}
where ${}_{2}\star$ denotes the Hodge dual on the two-sphere. The case $\ell_V = 1$ is special, as $\Grad_i \mathbb{Y}^{\ell_V,m_V}_j + \Grad_j \mathbb{Y}^{\ell_V,m_V}_i$ vanishes identically in this case, and will be treated separately later in the next subsection. For the scalar field we take
\begin{equation}
\delta\vec{\phi}(\tau,r,\theta,\varphi)=e^{-i\frac{2\pi\,n}{\beta\,L}\tau}\left[\sum_{I=1}^{\ell}\phi_{I}(r)(\underbrace{\Grad\ldots \Grad}_{I-1})_{i_1\ldots i_{I-1}}\mathbb{Y}_{i_I}^{m_V\,\ell_V}(\underbrace{\Grad\ldots \Grad}_{I})^{i_1\ldots i_{I}}\,\vec{X}\right]\,.
\label{eq:scavec}
\end{equation}
At this stage, our metric  and scalar field perturbations depend on a total of $\ell + 2$ variables.

We now turn our attention to infinitesimal diffeomorphisms. Within the symmetry class discussed in this subsection, the most general infinitesimal diffeomorphism is given by
\begin{equation}
\xi_a{\rm d}x^a=z(r){\rm d}y^i\mathbb{Y}_i^{m_V\,\ell_V}\,.
\label{eq:gauparamvector}
\end{equation}
Under such transformations, the metric and scalar field components transform as
\begin{align}
&\delta f_{r}(r)=z'(r)-\frac{2 r z(r)}{r^2+r_0^2}\,,\quad  \delta h_T(r)= z(r)\nonumber
\\
&
\delta\phi_1(r)=\frac{z(r)\phi(r)}{r^2+r_0^2}\,,\quad \text{and}\quad  \delta\phi_I(r)=0\,,\quad\text{for}\quad I\geq2\,.
\label{eq:gaugevec}
\end{align}

Because the action can be expressed entirely in terms of the Hamiltonian and momentum constraints, and since our metric deformations depend nontrivially only on the radial coordinate $r$, $f_r$ enters the quadratic action purely algebraically. Consequently, $f_r$ can be integrated out straightforwardly. As in the previous section, we assume an appropriate deformation of the integration contour to guarantee convergence of the quadratic path integral. This procedure is effectively equivalent to enforcing the constraint equations on off-shell configurations.

We now introduce $\ell$ gauge invariant variables
\begin{equation}
\begin{aligned}
&\phi_1(r)=\frac{\sqrt{2 \ell _V+1}}{2 \sqrt{2} \sqrt{\ell _V} \sqrt{\left(\ell _V+2\right) \left(\ell _V^2-1\right)}}\phi(r)\widehat{Q}_1(r)+\frac{h_T(r) \phi (r)}{r^2+r_0^2}\,,
\\
&\phi_I(r)=\frac{\sqrt{2 \ell _V+1}}{2 \sqrt{2} \sqrt{\ell _V} \sqrt{\left(\ell _V+2\right) \left(\ell _V^2-1\right)}}\widehat{Q}_I(r)\,,\quad \text{for}\quad I\in\{2,\ldots,\ell\}\,.
\end{aligned}
\end{equation}
Note that for backgrounds with $\ell=1$, we have a single $\widehat{Q}_I$ gauge invariant variable to consider. We now present explicit results for backgrounds with $\ell=1$ and $\ell=2$.

For a backgrounds with $\ell=1$ we find
\begin{equation}
\hat{S}^{[2]}=\frac{\beta}{16G}\int_{-\infty}^{+\infty}{\rm d}r\frac{\left(r^2+r_0^2\right) \sqrt{f(r)}}{\sqrt{g(r)}}\left[\frac{\phi (r)^2}{\left(\ell _V-1\right) \left(\ell _V+2\right)+4 \phi (r)^2}g(r)\left(\partial_rQ_1\right)^2+\frac{\phi (r)^2}{r^2+r_0^2}Q_1^2\right]\,,
\end{equation}
which is manifestly positive for any $\ell_V\geq2$.

For a background with $\ell=2$, we find
\begin{equation}
\hat{S}^{[2]}=\frac{\beta}{8G}\int_{-\infty}^{+\infty}{\rm d}r\frac{\left(r^2+r_0^2\right) \sqrt{f(r)}}{\sqrt{g(r)}}\left[H^{IJ}g(r)\left(\partial_rQ_I\right)\left(\partial_rQ_J\right)+V^{IJ}Q_IQ_J\right]
\end{equation}
with
\begin{equation}
H=\left[
\begin{array}{cc}
 \frac{6 \phi (r)^2}{12 \phi (r)^2+\left(\ell _V-1\right) \left(\ell _V+2\right)} & 0 \\
 0 & 6 \\
\end{array}
\right]\quad \text{and}\quad V= \left[
\begin{array}{cc}
 \frac{6 \phi (r)^2}{r^2+r_0^2} & -\frac{12 \phi (r)}{r^2+r_0^2} \\
 -\frac{12 \phi (r)}{r^2+r_0^2} & \frac{6 \omega ^2}{f(r)}+\frac{6 \left(\ell _V-1\right) \left(\ell _V+2\right)}{r^2+r_0^2} \\
\end{array}
\right]\,.
\end{equation}
$H$ is manifestly positive definite, and it is a simple exercise to show that $\det V > 0$ and ${\rm tr},V > 0$, thereby showing that $V$ is positive definite for $\ell_V \geq 2$, and thus that no negative modes exist in this sector.

\subsection{Static non-spherical vector-derived deformations with $\ell_V=1$}
This sector of perturbations still follows from Eq.(\ref{eq:vec}) and Eq.(\ref{eq:scavec}), but with $h_T$ effectively zero, since the combination $\nabla_i \mathbb{Y}^{\ell_V,m_V}_j + \nabla_j \mathbb{Y}^{\ell_V,m_V}_i$ vanishes identically. Linearized diffeomorphisms still induce the transformations given by Eq.~(\ref{eq:gaugevec}), but without $h_T$. Again, $f_r$ enters the action only algebraically and can therefore be readily integrated out. We are thus left with the $\phi_I$ fields alone. However, we note that for $\ell = 1$ there is a single $\phi_I$ to consider, and the same holds for $\ell \geq 2$, since
\begin{equation}
(\underbrace{\Grad\ldots \Grad}_{I-1})_{i_1\ldots i_{I-1}}\mathbb{Y}_{i_I}^{m_V\,\ell_V}(\underbrace{\Grad\ldots \Grad}_{I})^{i_1\ldots i_{I}}\,\vec{X}=0
\end{equation}
for $\ell_V = 1$ and $\ell \geq 2$. Finally, $\phi_1$ can be gauged away, leaving no negative modes in this sector (recall that $z\neq0$ in Eq.~(\ref{eq:gaugevec}) for $\ell_V=1$.). This, in turn, implies that these modes are the linearization of a pure-gauge mode in the full theory, as the method used by \cite{Alcubierre:2009ij} to show that Einstein-scalar theory admits a symmetric-hyperbolic formulation also applies to our system.

\subsection{Non-static and non-spherical scalar-derived deformations with $\ell_S\geq2$}

This sector builds upon its static counterpart. In many respects, the Ansatz closely resembles Eq.(\ref{eq:scalarstaticm}) and Eq.(\ref{eq:scalarstatics}), except that we need to accommodate for the dependence in $\tau$.

We thus take
\begin{subequations}
\begin{multline}
\delta {\rm d}s^2= \left[h_{\tau\tau}(\tau,r){\rm d}\tau^2+2h_{\tau r}(\tau,r){\rm d}\tau\,{\rm d}r+h_{rr}(\tau,r){\rm d}r^2+h_L(\tau,r){\rm d}\Omega_2^2\right]\mathbb{Y}^{m_S\,\ell_S}
\\
+2 h_{\tau}(\tau,r){\rm d}\tau\,{\rm d}y^i\Grad_i\mathbb{Y}^{m_S\,\ell_S}+2 h_r(\tau,r){\rm d}r\,{\rm d}y^i\Grad_i\mathbb{Y}^{m_S\,\ell_S}+h_T(\tau,r){\rm d}y^i\mathbb{Y}^{m_S\,\ell_S}_{ij}{\rm d}y^j\,,
\label{eq:scalarstaticmn}
\end{multline}
and
\begin{equation}
\delta\vec{\phi}(\tau,r,\theta,\varphi)=e^{-i\frac{2\pi\,n}{\beta\,L}\tau}\left[\phi_0(\tau,r)\vec{X}+\sum_{I=1}^{\ell}\phi_I(\tau,r)(\underbrace{\Grad\ldots \Grad}_{I})_{i_1\ldots i_I}\mathbb{Y}^{m_S\,\ell_S}(\underbrace{\Grad\ldots \Grad}_{I})^{i_1\ldots i_I}\,\vec{X}\right]\,.
\label{eq:scalarstaticsn}
\end{equation}
\end{subequations}
The full metric and scalar perturbations depend now on $\ell+8$ functions of $\tau$ and $r$. In contrast to the static sector, $\phi_{I}(\tau, r)$ and $\phi_{I}^\star(\tau, r)$ are treated as independent degrees of freedom. The $\tau$-dependence is then decomposed into Fourier modes:
\begin{equation}
\begin{aligned}
&h_{\tau\tau}(\tau,r)=\left[h_{\tau\tau,R}(r)+ih_{\tau\tau,I}(r)\right]e^{i \frac{2\pi k}{\beta\,L} \tau}+\left[h_{\tau\tau,R}(r)-ih_{\tau\tau,I}(r)\right]e^{-i \frac{2\pi k}{\beta\,L} \tau}\,,
\\
&h_{\tau r}(\tau,r)=\left[h_{\tau r,R}(r)+ih_{\tau r,I}(r)\right]e^{i \frac{2\pi k}{\beta\,L} \tau}+\left[h_{\tau r,R}(r)-ih_{\tau r,I}(r)\right]e^{-i \frac{2\pi k}{\beta\,L} \tau}\,,
\\
&h_{rr}(\tau,r)=\left[h_{rr,R}(r)+ih_{rr,I}(r)\right]e^{i \frac{2\pi k}{\beta\,L} \tau}+\left[h_{rr,R}(r)-ih_{rr,I}(r)\right]e^{-i \frac{2\pi k}{\beta\,L} \tau}\,,
\\
&h_{L}(\tau,r)=\left[h_{L,R}(r)+ih_{L,I}(r)\right]e^{i \frac{2\pi k}{\beta\,L} \tau}+\left[h_{L,R}(r)-ih_{L,I}(r)\right]e^{-i \frac{2\pi k}{\beta\,L} \tau}\,,
\\
&h_{\tau}(\tau,r)=\left[h_{\tau,R}(r)+ih_{\tau,I}(r)\right]e^{i \frac{2\pi k}{\beta\,L} \tau}+\left[h_{\tau,R}(r)-ih_{\tau,I}(r)\right]e^{-i \frac{2\pi k}{\beta\,L} \tau}\,,
\\
&h_{r}(\tau,r)=\left[h_{r,R}(r)+ih_{r,I}(r)\right]e^{i \frac{2\pi k}{\beta\,L} \tau}+\left[h_{r,R}(r)-ih_{r,I}(r)\right]e^{-i \frac{2\pi k}{\beta\,L} \tau}\,,
\\
&h_{T}(\tau,r)=\left[h_{T,R}(r)+ih_{T,I}(r)\right]e^{i \frac{2\pi k}{\beta\,L} \tau}+\left[h_{T,R}(r)-ih_{T,I}(r)\right]e^{-i \frac{2\pi k}{\beta\,L} \tau}\,,
\\
&\phi_I(\tau,r)=\left[\phi_{I,R}(r)+i\phi_{I,I}(r)\right]e^{i \frac{2\pi k}{\beta\,L} \tau}+\left[\hat{\phi}_{I,R}(r)-i\hat{\phi}_{I,I}(r)\right]e^{-i \frac{2\pi k}{\beta\,L} \tau}\,,
\\
&\phi_I^\star(\tau,r)=\left[\phi_{I,R}(r)-i\phi_{I,I}(r)\right]e^{-i \frac{2\pi k}{\beta\,L} \tau}+\left[\hat{\phi}_{I,R}(r)+i\hat{\phi}_{I,I}(r)\right]e^{i \frac{2\pi k}{\beta\,L} \tau}\,,
\end{aligned}
\end{equation}
where in the last two of the above $I=0,\ldots, \ell$. There are now a total of $4\ell+ 18$ variables, all depending only on $r$, that parametrize the most general deformation in this sector of perturbations. The most general infinitesimal diffeomorphism built out of spherical harmonics $\mathbb{Y}^{m_S\,\ell_S}$ depends now on three functions, namely
\begin{equation}
\xi_a{\rm d}x^a=\xi_{\tau}(\tau,r)\mathbb{Y}^{m_S\,\ell_S}\,{\rm d}\tau+\xi_{r}(\tau,r)\mathbb{Y}^{m_S\,\ell_S}\,{\rm d}r+z(\tau,r){\rm d}y^i\Grad_i\mathbb{Y}^{m_S\,\ell_S}\,,
\end{equation}
which we again decompose in terms of the following Fourier modes
\begin{equation}
\begin{aligned}
&\xi_{\tau}(\tau,r)=\left[\xi_{\tau,R}(r)+i\xi_{\tau,I}(r)\right]e^{i \frac{2\pi k}{\beta\,L} \tau}+\left[\xi_{\tau,R}(r)-i\xi_{\tau,I}(r)\right]e^{-i \frac{2\pi k}{\beta\,L} \tau}\,,
\\
&\xi_{r}(\tau,r)=\left[\xi_{r,R}(r)+i\xi_{r,I}(r)\right]e^{i \frac{2\pi k}{\beta\,L} \tau}+\left[\xi_{r,R}(r)-i\xi_{r,I}(r)\right]e^{-i \frac{2\pi k}{\beta\,L} \tau}\,,
\\
&z(\tau,r)=\left[z_{R}(r)+iz_{I}(r)\right]e^{i \frac{2\pi k}{\beta\,L} \tau}+\left[z_{R}(r)-iz_{I}(r)\right]e^{-i \frac{2\pi k}{\beta\,L} \tau}\,.
\end{aligned}
\end{equation}

Throughout the above, requiring regularity around the thermal circle imposes $k \in \mathbb{Z}$, with the special case $k = 0$ corresponding to the static sector.

Under such an infinitesimal diffeomorphism, the metric and scalar field deformations transform as
\begin{equation}
\begin{aligned}
&\delta h_{\tau\tau,R}(r)=g(r)f^\prime(r)\xi_{r,R}(r)-\frac{4\pi k}{\beta}\xi_{\tau,I}(r)\,,\quad \delta h_{\tau\tau,I}(r)=g(r)f^\prime(r)\xi_{r,I}(r)+\frac{4\pi k}{\beta}\xi_{\tau,R}(r)
\\
&\delta h_{\tau r,R}(r)=\xi_{\tau,R}^\prime(r)-\frac{f^\prime(r)}{f(r)}\xi_{\tau,R}(r)-\frac{2\pi k}{\beta}\xi_{r,I}(r)\,,\quad \delta h_{\tau r,I}(r)=\xi_{\tau,I}^\prime(r)-\frac{f^\prime(r)}{f(r)}\xi_{\tau,I}(r)+\frac{2\pi k}{\beta}\xi_{r,R}(r)
\\
&\delta h_{rr,R}(r)=2\xi_{r,R}^\prime(r)+\frac{g^\prime(r)}{g(r)}\xi_{r,R}(r)\,,\quad \delta h_{rr,I}(r)=2\xi_{r,I}^\prime(r)+\frac{g^\prime(r)}{g(r)}\xi_{r,I}(r)
\\
&\delta h_{\tau,R}(r)=\xi_{\tau,R}(r)-\frac{2\pi k}{\beta}z_I(r)\,,\quad \delta h_{\tau,I}(r)=\xi_{\tau,I}(r)+\frac{2\pi k}{\beta}z_R(r)
\\
&\delta h_{r,R}(r)=\xi_{r,R}(r)+z_R^\prime(r)-\frac{2rz_{R}(r)}{r^2+r_0^2}\,,\quad \delta h_{r,I}(r)=\xi_{r,I}(r)+z_I^\prime(r)-\frac{2rz_{I}(r)}{r^2+r_0^2}
\\
&\delta h_{L,R}(r)=-\ell_S(\ell_S+1)z_R(r)+2 r g(r)\xi_{r,R}(r)\,,\quad \delta h_{L,I}(r)=-\ell_S(\ell_S+1)z_I(r)+2 r g(r)\xi_{r,I}(r)\,,
\\
&\delta h_{T,R}(r)=2z_R(r)\,,\quad \delta h_{T,I}(r)=2z_I(r)\,,
\\
&\delta \phi_{0,R}(r)=\frac{\omega\,\xi_{\tau,I}(r)\phi(r)}{f(r)}+g(r)\xi_{r,R}(r)\phi^\prime(r)\,,\quad \delta \phi_{0,I}(r)=-\frac{\omega\,\xi_{\tau,R}(r)\phi(r)}{f(r)}+g(r)\xi_{r,I}(r)\phi^\prime(r)\,,
\\
&\delta \hat{\phi}_{0,R}(r)=-\frac{\omega\,\xi_{\tau,I}(r)\phi(r)}{f(r)}+g(r)\xi_{r,R}(r)\phi^\prime(r)\,,\quad \delta \hat{\phi}_{0,I}(r)=\frac{\omega\,\xi_{\tau,R}(r)\phi(r)}{f(r)}+g(r)\xi_{r,I}(r)\phi^\prime(r)\,,
\\
&\delta \phi_{1,R}(r)=\delta \hat{\phi}_{1,R}(r)=\frac{z_R(r)\phi(r)}{r^2+r_0^2}\,,\quad \delta \phi_{1,I}(r)=\delta \hat{\phi}_{1,I}(r)=\frac{z_I(r)\phi(r)}{r^2+r_0^2}\,,
\\
&\delta \phi_{I,R}(r)=\delta \phi_{I,I}(r)=\delta \hat{\phi}_{I,R}(r)=\delta \hat{\phi}_{I,I}(r)=0\quad\text{for}\quad I\geq2
\end{aligned}
\label{eq:gaugebonkers}
\end{equation}

Since the action is fully determined by the Hamiltonian and momentum constraints, and our metric deformations depend nontrivially only on the radial coordinate $r$, the fields $h_{\tau r,R}$, $h_{\tau r,I}$, $h_{r r,R}$, $h_{r r,I}$, $f_{r,R}$ and $f_{r,I}$ enter the quadratic action purely algebraically. Consequently, they can be readily integrated out. We are thus left with $4\ell+12$ variables to control. However, these are still gauge dependent, and thus to proceed we introduce gauge invariant variables
\begin{equation}
\begin{aligned}
& Q_{1,R}(r)=h_{\tau \tau,R}(r)- \left[\frac{\ell _S \left(\ell _S+1\right)f'(r)}{4 r}+\frac{4 \pi ^2 k^2}{\beta ^2}\right]h_{T,R}(r)-\frac{f'(r) h_{L,R}(r)}{2 r}+\frac{4 \pi  k h_{\tau ,I}(r)}{\beta }\,,
\\
& Q_{2,R}(r)=\phi _{0,R}(r)+\left[\frac{\pi  k \omega  \phi (r)}{\beta f(r)}-\frac{\ell _S \left(\ell _S+1\right) \phi '(r)}{4 r}\right]h_{T,R}(r) -\frac{\omega  \phi (r) h_{\tau ,I}(r)}{f(r)}-\frac{\phi '(r) \
h_{L,R}(r)}{2 r}\,,
\\
& Q_{3,R}(r)=\hat{\phi }_{0,R}(r)- \left[\frac{\pi  k \omega  \phi (r)}{\beta f(r)}+\frac{\ell _S \left(\ell _S+1\right) \phi '(r)}{4 r}\right]h_{T,R}(r)+\frac{\omega  \phi (r) h_{\tau ,I}(r)}{f(r)}-\frac{\phi '(r) h_{L,R}(r)}{2 r}\,,
\\
& Q_{4,R}(r)=\phi _{1,R}(r)-\frac{\phi (r) h_{T,R}(r)}{2 \
\left(r^2+r_0^2\right)}\,,
\\
& Q_{5,R}(r)=\hat{\phi }_{1,R}(r)-\frac{\phi (r) h_{T,R}(r)}{2 \left(r^2+r_0^2\right)}\,,
\\
& Q_{2I+2,R}(r)=\phi _{I,R}(r)\,,\quad\text{for}\quad  I\geq 2\,,
\\
& Q_{2I+3,R}(r)=\hat{\phi} _{I,R}(r)\,,\quad\text{for}\quad  I\geq 2\,,
\end{aligned}
\end{equation}
and similarly for the $I$ sector
\begin{equation}
\begin{aligned}
 &Q_{1,I}(r)=h_{\tau \tau,I}(r)-\left[\frac{\ell _S \left(\ell _S+1\right) f'(r)}{4 r}+\frac{4 \pi ^2 k^2}{\beta ^2}\right]h_{T,I}(r) -\frac{f'(r) h_{L,I}(r)}{2 r}-\frac{4 \pi  k h_{\tau ,R}(r)}{\beta }\,,
 \\
& Q_{2,I}(r)=\phi _{0,I}(r)+\left[\frac{\pi  k \omega  \phi (r)}{\beta  f(r)}-\frac{\ell _S \left(\ell _S+1\right) \phi '(r)}{4 r}\right]h_{T,I}(r) +\frac{\omega  \phi (r) h_{\tau ,R}(r)}{f(r)}-\frac{\phi '(r)h_{L,I}(r)}{2 r}\,,
\\
& Q_{3,I}(r)=\hat{\phi }_{0,I}(r) - \left[\frac{\pi  k \omega  \phi (r)}{\beta f(r)}+\frac{\ell _S \left(\ell _S+1\right) \phi '(r)}{4 r}\right]h_{T,I}(r)-\frac{\omega  \phi (r) h_{\tau ,R}(r)}{f(r)}-\frac{\phi '(r) h_{L,I}(r)}{2 r}\,,
\\
& Q_{4,I}(r)=\phi _{1,I}(r)-\frac{\phi (r) h_{T,I}(r)}{2 \
\left(r^2+r_0^2\right)}\,,
\\
& Q_{5,I}(r)=\hat{\phi }_{1,I}(r)-\frac{\phi (r) h_{T,I}(r)}{2 \
\left(r^2+r_0^2\right)}\,,
\\
& Q_{2I+2,I}(r)=\phi _{I,I}(r)\,,\quad\text{for}\quad  I\geq 2\,,
\\
& Q_{2I+3,I}(r)=\hat{\phi} _{I,I}(r)\,,\quad\text{for}\quad  I\geq 2\,,
\end{aligned}
\end{equation}

After several integrations by parts, the quadratic action ultimately depends on just $4\ell + 6$ variables, namely the components of $Q_I(r)$: its real and imaginary parts. Furthermore, due to the $U(1)$ symmetry of the background around the Euclidean time circle, the $R$ and $I$ sectors are isopectral and share identical quadratic actions. We may therefore, without loss of generality, focus on the $R$ sector, which involves $2\ell + 3$ arbitrary functions. This is the sector of perturbations that we investigate numerically.

\subsection{\label{app:D7}Non-static and non-spherical scalar-derived deformations with $\ell_S=1$}
We now turn our attention to the final scalar sector. This sector of deformations is not very different from the previous one, except that the term multiplying $h_T(\tau,r)$ in the metric deformation given in Eq.(\ref{eq:scalarstaticmn}) vanishes. This has little consequence, except when we need to construct gauge-invariant variables. In particular, under a generic diffeomorphism generated by an $\ell_S=1$ deformation, all variables still transform as in Eq.(\ref{eq:gaugebonkers}) with $\ell_S=1$, and without involving $h_{T,R}$ or $h_{T,I}$. Additionally, we can still integrate out $h_{\tau r,R}$, $h_{\tau r,I}$, $h_{r r,R}$, $h_{r r,I}$, $f_{r,R}$, and $f_{r,I}$, leaving a quadratic action that depends on $4\ell + 10$ remaining variables to control.

At this stage we introduce new gauge invariant variables
\begin{equation}
\begin{aligned}
 &Q_{1,R}(r)=\phi _{0,R}(r)-\frac{\pi  k t_+(r) h_{L,R}(r)}{f(r) z(r)}+\frac{\beta  z_-(r) h_{\tau \tau ,R}(r)}{f(r) z(r)}-\frac{\beta  t_+(r) h_{\tau ,I}(r)}{f(r) z(r)}\,,
\\
 &Q_{2,R}(r)=\hat{\phi}_{0,R}(r)+\frac{\pi  k t_-(r) h_{L,R}(r)}{f(r) z(r)}-\frac{\beta  z_+(r) h_{\tau \tau ,R}(r)}{f(r) z(r)}+\frac{\beta  t_-(r) h_{\tau ,I}(r)}{f(r) z(r)}\,,
\\
 &Q_{3,R}(r)=\phi _{1,R}(r)+\frac{\beta ^2 \phi (r) f'(r) h_{L,R}(r)}{2 \left(r^2+r_0^2\right) z(r)}-\frac{4 \pi  \beta  k r \phi (r) h_{\tau,I}(r)}{\left(r^2+r_0^2\right) z(r)}-\frac{\beta ^2 r \phi (r) h_{\tau \tau ,R}(r)}{\left(r^2+r_0^2\right) z(r)}\,,
\\
 &Q_{4,R}(r)=\hat{\phi }_{1,R}(r)+\frac{\beta ^2 \phi (r) f'(r) h_{L,R}(r)}{2 \
\left(r^2+r_0^2\right) z(r)}-\frac{4 \pi  \beta  k r \phi (r) h_{\tau,I}(r)}{\left(r^2+r_0^2\right) z(r)}-\frac{\beta ^2 r \phi (r) h_{\tau \tau ,R}(r)}{\left(r^2+r_0^2\right) z(r)}\,,
\\
& Q_{2I+1,R}(r)=\phi _{I,R}(r)\,,\quad\text{for}\quad  I\geq 2\,,
\\
& Q_{2I+2,R}(r)=\hat{\phi} _{I,R}(r)\,,\quad\text{for}\quad  I\geq 2\,,
\end{aligned}
\end{equation}
and similarly for the $I$ sector
\begin{equation}
\begin{aligned}
&  Q_{1,I}(r)=\phi _{0,I}(r)-\frac{\pi  k t_+(r) h_{L,I}(r)}{f(r) z(r)}+\frac{\beta  t_+(r) h_{\tau ,R}(r)}{f(r) z(r)}+\frac{\beta  z_-(r) h_{\tau \tau ,I}(r)}{f(r) z(r)}\,,
\\
& Q_{2,I}(r)=\hat{\phi }_{0,I}(r)-\frac{\beta  t_-(r)h_{\tau ,R}(r) }{f(r) z(r)}+\frac{\pi  k \
t_-(r) h_{L,I}(r)}{f(r) z(r)}-\frac{\beta  z_+(r) h_{\tau \tau ,I}(r)}{f(r) z(r)}\,,
\\
& Q_{3,I}(r)=\phi _{1,I}(r)+\frac{\beta ^2 \phi (r) f'(r) h_{L,I}(r)}{2\left(r^2+r_0^2\right) z(r)}+\frac{4 \pi  \beta  k r \phi (r) h_{\tau,R}(r)}{\left(r^2+r_0^2\right) z(r)}-\frac{\beta ^2 r \phi (r) h_{\tau \tau ,I}(r)}{\left(r^2+r_0^2\right) z(r)}\,,
\\
& Q_{4,I}(r)=\hat{\phi }_{1,I}(r)+\frac{\beta ^2 \phi (r) f'(r) h_{L,I}(r)}{2 \
\left(r^2+r_0^2\right) z(r)}+\frac{4 \pi  \beta  k r \phi (r) h_{\tau,R}(r)}{\left(r^2+r_0^2\right) z(r)}-\frac{\beta ^2 r \phi (r) h_{\tau \tau ,I}(r)}{\left(r^2+r_0^2\right) z(r)}\,,
\\
& Q_{2I+1,I}(r)=\phi _{I,R}(r)\,,\quad\text{for}\quad  I\geq 2\,,
\\
& Q_{2I+2,I}(r)=\hat{\phi} _{I,R}(r)\,,\quad\text{for}\quad  I\geq 2\,,
\end{aligned}
\end{equation}
where
\begin{equation}
\begin{aligned}
&z(r)=\beta ^2 f'(r)+8 \pi ^2 k^2 r\,,
\\
&z_{\pm}(r)=\beta  f(r) \phi '(r)\pm2 \pi  k r \omega  \phi (r)\,,
\\
&t_{\pm}(r)=\beta  \omega  \phi (r) f'(r)\pm4 \pi  k f(r) \phi '(r)\,,
\end{aligned}
\end{equation}
After several integrations by parts, the quadratic action ultimately depends on just $4\ell + 4$ variables, namely $\{Q_{I\,R}(r),Q_{I\,I}(r)\}$. Once again, due to the $U(1)$ symmetry around the background thermal circle, the $R$ and $I$ sectors decouple and are isopectral. We can therefore focus on just one of them—say, the $R$ sector—resulting in a quadratic action that depends on $2\ell + 2$ variables, which we analyze numerically.

\begin{figure}
    \centering
    \includegraphics[width=\linewidth]{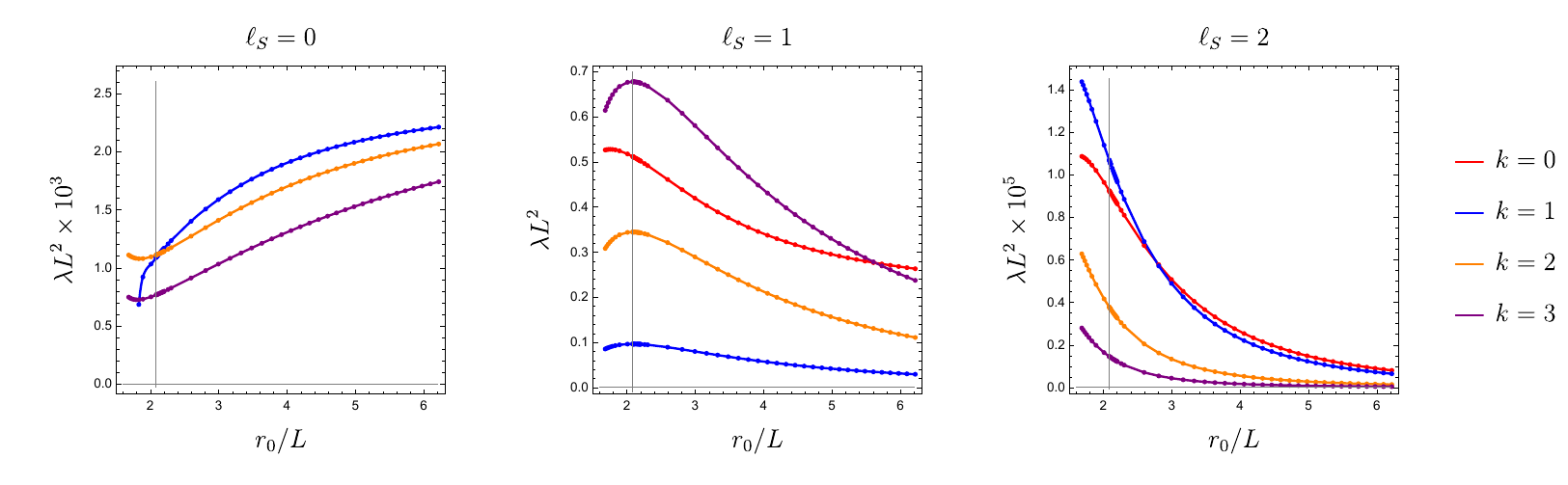}
    \caption{The lowest lying mode of $\mathcal{L}$ in the even sector for wormholes with $\ell=1,n=\beta=1$.   The horizontal lines are $\lambda=0$ while the vertical lines mark the critical value of $r_0$ from figure~\ref{fig:fxibeta}. There are no negative modes in these sectors.}
    \label{fig:higherspec}
\end{figure}

\subsection{Non-static and non-spherical vector-derived deformations with $\ell_V\geq2$}
The Ansatz for this sector of perturbations extends the static case by incorporating deformations that depend on $\tau$. Although this is the second most intricate sector, it turns out to be relatively straightforward to analyze.

Our metric and scalar field deformations read
\begin{subequations}
\begin{equation}
\delta {\rm d}s^2=2f_r(\tau,r)\,\mathbb{Y}^{\ell_V\,m_V}_{i}\,{\rm d}r\,{\rm d}y^i+h_T(\tau,r)\left(\Grad_i\mathbb{Y}^{\ell_V\,m_V}_j+\Grad_j\mathbb{Y}^{\ell_V\,m_V}_i\right)\,{\rm d}y^i{\rm d}y^j
\label{eq:vectau}
\end{equation}
and
\begin{equation}
\delta\vec{\phi}(\tau,r,\theta,\varphi)=e^{-i\frac{2\pi\,n}{\beta\,L}\tau}\left[\sum_{I=1}^{\ell}\phi_{I}(\tau,r)(\underbrace{\Grad\ldots \Grad}_{I-1})_{i_1\ldots i_{I-1}}\mathbb{Y}_{i_I}^{m_V\,\ell_V}(\underbrace{\Grad\ldots \Grad}_{I})^{i_1\ldots i_{I}}\,\vec{X}\right]\,,
\label{eq:scavectau}
\end{equation}
\end{subequations}
and similarly for $\delta\vec{\phi^\star}$. Unlike in the static sector, $\phi_{I}(\tau,r)$ and $\phi_{I}^\star(\tau,r)$ are to be regarded as independent degrees of freedom. We further decompose the dependence on $\tau$ in terms of Fourier modes:
\begin{equation}
\begin{aligned}
&f_{\tau}(\tau,r)=\left[f_{\tau,R}(r)+if_{\tau,I}(r)\right]e^{i \frac{2\pi k}{\beta\,L} \tau}+\left[f_{\tau,R}(r)-if_{\tau,I}(r)\right]e^{-i \frac{2\pi k}{\beta\,L} \tau}
\\
&f_{r}(\tau,r)=\left[f_{r,R}(r)+if_{r,I}(r)\right]e^{i \frac{2\pi k}{\beta\,L} \tau}+\left[f_{r,R}(r)-if_{r,I}(r)\right]e^{-i \frac{2\pi k}{\beta\,L} \tau}
\\
&h_{T}(\tau,r)=\left[h_{T,R}(r)+ih_{T,I}(r)\right]e^{i \frac{2\pi k}{\beta\,L} \tau}+\left[h_{T,R}(r)-ih_{T,I}(r)\right]e^{-i \frac{2\pi k}{\beta\,L} \tau}
\\
&\phi_I(\tau,r)=\left[\phi_{I,R}(r)+i\phi_{I,I}(r)\right]e^{i \frac{2\pi k}{\beta\,L} \tau}+\left[\hat{\phi}_{I,R}(r)-i\hat{\phi}_{I,I}(r)\right]e^{-i \frac{2\pi k}{\beta\,L} \tau}
\\
&\phi_I^\star(\tau,r)=\left[\phi_{I,R}(r)-i\phi_{I,I}(r)\right]e^{-i \frac{2\pi k}{\beta\,L} \tau}+\left[\hat{\phi}_{I,R}(r)+i\hat{\phi}_{I,I}(r)\right]e^{i \frac{2\pi k}{\beta\,L} \tau}\,.
\end{aligned}
\end{equation}
There are a total of $4 + 4\ell$ variables that depend only on $r$, which we need to analyze. Infinitesimal diffeomorphisms admit a similar decomposition as
\begin{subequations}
\begin{equation}
\xi_a{\rm d}x^a=z(\tau,r){\rm d}y^i\mathbb{Y}_i^{m_V\,\ell_V}
\label{eq:gauparamvectortau}
\end{equation}
with
\begin{equation}
z(\tau,r)=\left[z_{R}(r)+iz_{I}(r)\right]e^{i \frac{2\pi k}{\beta\,L} \tau}+\left[z_{R}(r)-iz_{I}(r)\right]e^{-i \frac{2\pi k}{\beta\,L} \tau}\,.
\end{equation}
\end{subequations}
In all of the above, regularity of the deformations around the thermal circle, demands $k\in\mathbb{Z}$, with $k=0$ yielding back the static sector. Note also that, by construction, all the functions that depend on $r$ only are real. It is a simple exercise to show how each of the metric and scalar field deformations transform under the generic diffeomorphisms above:
\begin{equation}
\begin{aligned}
&\delta f_{\tau,R}(r)=-\frac{2\pi k}{\beta}z_I(r)\,,\quad \delta f_{\tau,I}(r)=\frac{2\pi k}{\beta}z_R(r)\,,\quad \delta f_{r,R}(r)=z_R^\prime(r)-\frac{2\,r\,z_R(r)}{r^2+r_0^2}\,,
\\
&\delta f_{r,I}(r)=z_I^\prime(r)-\frac{2\,r\,z_I(r)}{r^2+r_0^2}\,,\quad \delta h_{T,R}(r)=z_R(r)\,,\quad \delta h_{T,I}(r)=z_I(r)\,,
\\
&\delta \phi_{1,R}(r)=\delta \hat{\phi}_{1,R}(r)=\frac{z_R(r)\phi(r)}{r^2+r_0^2}\,,\quad \delta \phi_{1,I}(r)=\delta \hat{\phi}_{1,I}(r)=\frac{z_I(r)\phi(r)}{r^2+r_0^2}\,,
\\
&\delta \phi_{I,R}(r)=\delta \phi_{I,I}(r)=\delta \hat{\phi}_{I,R}(r)=\delta \hat{\phi}_{I,I}(r)=0\quad\text{for}\quad I\geq2\,.
\end{aligned}
\end{equation}

Since the action is fully determined by the Hamiltonian and momentum constraints, and our metric deformations vary nontrivially only with the radial coordinate $r$, the fields $f_{r,R}$ and $f_{r,I}$ appear in the quadratic action in an entirely algebraic manner. As a result, $f_{r,R}$ and $f_{r,I}$ can be readily integrated out. A this stage we have a quadratic action that depends on $f_{\tau,R}$, $f_{\tau,I}$, $\phi_{I,R}$,  $\phi_{I,I}$, $\hat{\phi}_{I,R}$ and $\phi_{I,I}$. We now introduce gauge invariant variables as follows
\begin{equation}
\begin{aligned}
&\hat{Q}_{1,R}(r)=f_{\tau,I}(r)-\frac{2\pi k}{\beta}h_{T,R}(r)\,,\quad \hat{Q}_{1,I}(r)=f_{\tau,R}(r)+\frac{2\pi k}{\beta}h_{T,I}(r)\,,
\\
&\hat{Q}_{2,R}(r)=\phi_{1,R}(r)-\frac{\phi(r)}{r^2+r_0^2}h_{T,R}(r)\,,\quad \hat{Q}_{2,I}(r)=\phi_{1,I}(r)-\frac{\phi(r)}{r^2+r_0^2}h_{T,I}(r)\,,
\\
&\hat{Q}_{3,R}(r)=\hat{\phi}_{1,R}(r)-\frac{\phi(r)}{r^2+r_0^2}h_{T,R}(r)\,,\quad \hat{Q}_{3,I}(r)=\hat{\phi}_{1,I}(r)-\frac{\phi(r)}{r^2+r_0^2}h_{T,I}(r)\,,
\\
&\hat{Q}_{2I,R}(r)=\phi_{I,R}(r)\,,\quad \hat{Q}_{2I,I}(r)=\phi_{I,I}(r)\,,\quad\text{for}\quad I\geq2\,,
\\
&\hat{Q}_{2I+1,R}(r)=\hat{\phi}_{I,R}(r)\,,\quad \hat{Q}_{2I+1,I}(r)=\hat{\phi}_{I,I}(r)\,,\quad\text{for}\quad I\geq2\,.
\end{aligned}
\end{equation}
The quadratic action now depends on $4\ell + 2$ gauge-invariant quantities, $\{\hat{Q}_{\hat{I},R}, \hat{Q}_{\hat{I},I}\}$, with $
\hat{I}=1,\ldots,4\ell+2$. The $R$ and $I$ sectors decouple from each other and share identical quadratic actions; in other words, they are isospectral. This is expected, as they are related by the action of the $U(1)_\tau$ symmetry, which maps the relevant sine functions to cosine functions. We are thus left with $2\ell+1$ variables to study, which we can pick as the $R$ sector. It turns out that the quadratic action is best written in terms of some simple variables that relate to the $\hat{Q}_{\hat{I},R}$
\begin{equation}
\begin{aligned}
&\hat{Q}_{1,R}(r)=(r^2+r_0^2)\left[\frac{Q_1(r)}{L}-\frac{2\pi k}{\beta}\sum_{\hat{I}=1}^{\ell}Q_{2\hat{I}+1}(r)\right]\,,
\\
&\hat{Q}_{2\hat{I},R}(r)=Q_{2\hat{I}}(r)+\phi(r)Q_{2\hat{I}+1}(r)\,,
\\
&\hat{Q}_{2\hat{I}+1,R}(r)=-Q_{2\hat{I}}(r)+\phi(r)Q_{2\hat{I}+1}(r)\,,
\end{aligned}
\end{equation}
with $\hat{I}=1,\ldots,\ell$.

The relevant quadratic action, up to multiplicative positive constants, takes the following form
\begin{subequations}
\begin{equation}
\hat{S}^{[2]}\supset\int_{-\infty}^{+\infty}{\rm d}r\frac{\left(r^2+r_0^2\right) \sqrt{f(r)}}{\sqrt{g(r)}}\left[g(r)H^{\hat{I}\hat{J}}\partial_rQ_{\hat{I}}\partial_rQ_{\hat{J}}+V^{\hat{I}\hat{J}}Q_{\hat{I}}Q_{\hat{J}}\right]\,.
\end{equation}
We investigated the positivity properties of $H^{\hat{I}\hat{J}}$ and $V^{\hat{I}\hat{J}}$ for wormholes with $\ell = 1$, $\ell = 2$, $\beta \in (2 \times 10^{-2}, 10)$, and for $V \in (V_{\min}, 10)$, and found both matrices to be positive definite for any value of $\ell_V \geq 2$ and $k \in \mathbb{Z}$. For completeness, we now present explicit expressions for $H^{\hat{I}\hat{J}}$ and $V^{\hat{I}\hat{J}}$ in the case of $\ell = 1$ wormholes:
\begin{equation}
H=\left[
\begin{array}{ccc}
 \frac{4 \beta ^3 \left(r^2+r_0^2\right) \ell _V \left(\ell _V+1\right) \left(4 \phi (r)^2+\ell _V^2+\ell _V-2\right)}{L^2 \hat{z}(r) \left(2 \ell _V+1\right)} & 0 & -\frac{8 \pi  \beta ^2 k 
   \left(r^2+r_0^2\right) \ell _V \left(\ell _V^3+2 \ell _V^2-\ell _V-2\right)}{L \hat{z}(r) \left(2 \ell _V+1\right)} \\
 0 & \frac{16 \beta  \ell _V \left(\ell _V+1\right)}{2 \ell _V+1} & 0 \\
 -\frac{8 \pi  \beta ^2 k  \left(r^2+r_0^2\right) \ell _V \left(\ell _V^3+2 \ell _V^2-\ell _V-2\right)}{L \hat{z}(r) \left(2 \ell _V+1\right)} & 0 & \frac{16 \beta  \check{z}(r) \ell _V \left(\ell _V^3+2
   \ell _V^2-\ell _V-2\right)}{\hat{z}(r) \left(2 \ell _V+1\right)} \\
\end{array}
\right]\,,
\end{equation}
and
\begin{equation}
V=\left[
\begin{array}{ccc}
 \frac{4 \beta  \ell _V \left(\ell _V+1\right) \left(4 \phi (r)^2+\ell _V^2+\ell _V-2\right)}{L^2 f(r) \left(2 \ell _V+1\right)} & -\frac{32 \beta  \omega  \phi (r) \ell _V \left(\ell _V+1\right)}{f(r)
   L\left(2\ell _V+1\right)} & -\frac{8 \pi  k  \ell _V \left(\ell _V^3+2 \ell _V^2-\ell _V-2\right)}{f(r) L\left(2 \ell _V+1\right)} \\
 -\frac{32 \beta  \omega  \phi (r) \ell _V \left(\ell _V+1\right)}{f(r) L\left(2\ell _V+1\right)} & \frac{16 \ell _V \left(\ell _V+1\right) \tilde{z}(r)}{\left(r^2+r_0^2\right) f(r) \beta \left(2\ell _V+1\right)} & 0 \\
 -\frac{8 \pi  k  \ell _V \left(\ell _V^3+2 \ell _V^2-\ell _V-2\right)}{f(r) L\left(2 \ell _V+1\right)} & 0 & \frac{16 \check{z}(r) \ell _V \left(\ell _V^3+2 \ell _V^2-\ell
   _V-2\right)}{\left(r^2+r_0^2\right) f(r) \beta \left(2\ell_V+1\right)} \\
\end{array}
\right]\,,
\end{equation}
where
\begin{equation}
\begin{aligned}
&\hat{z}(r)=\beta ^2 f(r) \left[4 \phi (r)^2+\ell _V^2+\ell _V-2\right]+4 \pi ^2 k^2 \left(r^2+r_0^2\right)\,,
\\
&\check{z}(r)=\beta ^2 f(r) \phi (r)^2+\pi^2 k^2 \left(r^2+r_0^2\right)\,,
\\
&\tilde{z}(r)=\beta ^2 f(r) \ell _V \left(\ell _V+1\right)+\left(r^2+r_0^2\right) \left(\beta ^2 \omega ^2+4 \pi ^2 k^2\right)\,.
\end{aligned}
\end{equation}
\end{subequations}

\subsection{Non-static and non-spherical vector-derived deformations with $\ell_V=1$}
Lastly, we study vector-derived deformations that depend on the Euclidean circle, with $\ell_V = 1$. In many ways, this sector resembles the previous case, except that $h_T$ does not appear in the ansatz for the metric deformations (see Eq.~(\ref{eq:vectau})). As a result, we must construct different gauge-invariant variables. Up to the point where gauge invariant variables are introduced, the quadratic action can simply be obtained by taking the limit $\ell_V \to 1$ in the results of the previous section.

The new gauge invariant variables read
\begin{equation}
\begin{aligned}
&\hat{Q}_{1,R}(r)=\phi_{1,R}(r)-\frac{\beta\phi(r)}{2\pi k(r^2+r_0^2)}f_{\tau,I}(r)\,,\quad \hat{Q}_{1,I}(r)=\phi_{1,I}(r)+\frac{\beta\phi(r)}{2\pi k(r^2+r_0^2)}f_{\tau,R}(r)\,,
\\
&\hat{Q}_{2,R}(r)=\hat{\phi}_{1,R}(r)-\frac{\beta\phi(r)}{2\pi k(r^2+r_0^2)}f_{\tau,I}(r)\,,\quad \hat{Q}_{2,I}(r)=\hat{\phi}_{1,I}(r)+\frac{\beta \phi(r)}{2\pi k(r^2+r_0^2)}f_{\tau,R}(r)\,,
\\
&\hat{Q}_{2I-1,R}(r)=\phi_{I,R}(r)\,,\quad \hat{Q}_{2I-1,I}(r)=\phi_{I,I}(r)\,,\quad\text{for}\quad I\geq2\,,
\\
&\hat{Q}_{2I,R}(r)=\hat{\phi}_{I,R}(r)\,,\quad \hat{Q}_{2I,I}(r)=\hat{\phi}_{I,I}(r)\,,\quad\text{for}\quad I\geq2\,.
\end{aligned}
\end{equation}

Once again, the $R$ and $I$ sectors decouple from each other and share identical quadratic actions; in other words, they are isospectral. This is expected, as they are related by the action of the $U(1)\tau$ symmetry, which maps the relevant sine functions to cosine functions. As a result, we are left with $2\ell$ variables to study, which we choose to take from the $R$ sector. It turns out that the quadratic action is most naturally expressed in terms of a set of simple variables related to the $\hat{Q}{\hat{I},R}$, defined as
\begin{equation}
\begin{aligned}
&\hat{Q}_{2\hat{I}-1,R}(r) = Q_{2\hat{I}-1}(r) + \phi(r) Q_{2\hat{I}}(r) \,,
\\
&\hat{Q}_{2\hat{I},R}(r) = -Q_{2\hat{I}-1}(r) + \phi(r) Q_{2\hat{I}}(r) 
\end{aligned}
\end{equation}
with $\hat{I} = 1, \ldots, \ell$.

Up to an overall positive constant, the relevant quadratic action takes the following form:
\begin{subequations}
\begin{equation}
\hat{S}^{[2]}\supset\int_{-\infty}^{+\infty}{\rm d}r\frac{\left(r^2+r_0^2\right) \sqrt{f(r)}}{\sqrt{g(r)}}\left[g(r)H^{\hat{I}\hat{J}}\partial_rQ_{\hat{I}}\partial_rQ_{\hat{J}}+V^{\hat{I}\hat{J}}Q_{\hat{I}}Q_{\hat{J}}\right]\,.
\end{equation}
As in previous cases, we examined the positivity properties of $H^{\hat{I}\hat{J}}$ and $V^{\hat{I}\hat{J}}$ for wormholes with $\ell = 1$ and $\ell = 2$, across the range $\beta \in (2 \times 10^{-2}, 10)$ and for $V \in (V_{\min}, 10)$. Our analysis showed that both matrices remain positive definite for all values of $k \in \mathbb{Z}$. For completeness, we now provide the explicit forms of $H^{\hat{I}\hat{J}}$ and $V^{\hat{I}\hat{J}}$ in the specific case of $\ell = 1$ wormholes:
\begin{equation}
H=\left[
\begin{array}{cc}
 \frac{32 \beta }{3} & 0 \\
 0 & \frac{32 \pi ^2 \beta  k^2 \left(r^2+r_0^2\right) \phi (r)^2}{3 \check{z}(r)} \\
\end{array}
\right]\,,
\end{equation}
and
\begin{equation}
V=\left[
\begin{array}{cc}
 \frac{32 \left(\beta ^2 \omega ^2+4 \pi ^2 k^2\right)}{3 \beta  f(r)}+\frac{64 \beta }{3 \left(r^2+r_0^2\right)} & -\frac{128 \pi  k \omega  \phi (r)}{3 f(r)} \\
 -\frac{128 \pi  k \omega  \phi (r)}{3 f(r)} & \frac{128 \pi ^2 k^2 \phi (r)^2}{3 \beta  f(r)} \\
\end{array}
\right]\,.
\end{equation}
\end{subequations}
It is a simple exercise to check that both $H$ and $V$ defined above are positive definite for $\phi(r)\neq0$.

\section{The operator approach}

For all sectors of perturbations, the quadratic action expressed in terms of gauge-invariant variables takes the form
\begin{subequations}
\begin{equation}
S^{[2]} \propto \int_{\mathcal{M}} {\rm d}^4x \, \sqrt{g} \left[
\mathcal{K}^{IJ} g^{ab} \mathcal{D}_a\psi_{I} \mathcal{D}_b\psi_{J}
+ \tilde{V}^{IJ} \psi_I \psi_J
\right] ,
\end{equation}
with
\begin{equation}
\mathcal{D}_a\psi_{I} = \nabla_a \psi_I + A_{a \phantom{J} I}^{\phantom{a}J} \psi_J \, .
\end{equation}
\end{subequations}
Here $\mathcal{K}^{IJ}$, $V^{IJ}$, and $A_{a\phantom{J}I}^{\phantom{a}J}$ are spacetime-dependent tensors. Defining $A_{aIJ} \equiv \mathcal{K}_{IK} A_{a\phantom{K}J}^{\phantom{a}K}$, one finds that $A_{aIJ}$ is antisymmetric under the exchange $I \leftrightarrow J$. Obtaining this form involves performing multiple integrations by parts. Nevertheless, the boundary terms generated in the process vanish due to our chosen boundary conditions. For all cases under consideration, $\mathcal{K}^{IJ}$ is positive definite throughout the large wormhole branch and across much of the small wormhole branch, while $\tilde{V}^{IJ}$ fails to maintain this property.

Upon integration by parts, one finds
\begin{subequations}
\begin{equation}
S^{[2]} \propto \int_{\mathcal{M}} {\rm d}^4x \, \sqrt{g} \psi_I (\mathcal{L}\psi)^I ,
\end{equation}
with
\begin{equation}
(\mathcal{L}\psi)^I
\equiv - \mathcal{D}_a\!\left[\mathcal{K}^{IJ} g^{ab} \mathcal{D}_b\psi_{J}\right]
+ \tilde{V}^{IJ}\psi_J ,
\qquad
\mathcal{D}_a X^I \equiv \nabla_a X^I +A_{a}^{\phantom{a}IJ} X_J .
\end{equation}
\end{subequations}
One may then search for negative modes by solving the eigenvalue problem
\begin{equation}
- \mathcal{D}_a\!\left(\mathcal{K}^{IJ} g^{ab} \mathcal{D}_b \psi_{J}\right)
+ \tilde{V}^{IJ}\psi_J \;=\; \lambda \,\mathcal{K}^{IJ}\psi_J \,.
\end{equation}
We have applied this approach in parallel with the action-based method, and found complete agreement in the number of negative modes present in each sector. As expected, the numerical values of the eigenvalues~$\lambda$ differ, reflecting the distinct choices of inner product used in the two formulations. In \cite{Marolf:2021kjc}, this method was applied extensively to examine whether negative modes arise across a wide variety of wormhole solutions in both gravity and supergravity theories.

\bibliographystyle{JHEP}
\bibliography{refs}

\end{document}